\documentclass[useAMS,usenatbib,usegraphicx]{mn2e}

\newcommand{\vt}{\mbox{\bf {T}}}
\newcommand{\vC}{\mbox{\bf{C}}}
\newcommand{\vm}{\mbox{\bf {M}}}

\newcommand{\vx}{\mbox{\bf {x}}}

\newcommand{\vk}{\mbox{\bf {k}}}

\newcommand{\vq}{\mbox{\bf {q}}}

\newcommand{\vrv}{\mbox{\bf {r}}}

\newcommand{\n}{\hat{\mbox{\bf {n}}}}
\newcommand{\kh}{\hat{\mbox{\bf {k}}}}
\newcommand{\qh}{\hat{\mbox{\bf {q}}}}

\newcommand{\vv}{\mbox{\bf {v}}}

\newcommand{\dthreek}{\frac{d\vk}{(2\pi)^3}}
\newcommand{\dthreekuno}{\frac{d\vk_1}{(2\pi)^3}}
\newcommand{\dthreekdos}{\frac{d\vk_2}{(2\pi)^3}}
\newcommand{\dthreeq}{\frac{d\vq}{(2\pi)^3}}
\newcommand{\dthreequno}{\frac{d\vq_1}{(2\pi)^3}}
\newcommand{\dthreeqdos}{\frac{d\vq_2}{(2\pi)^3}}

\input epsf

\def\plotancho#1{\includegraphics[width=18cm]{#1}}

\title{On the peculiar momentum of baryons after Reionization}

\author[C. Hern\'andez--Monteagudo \& S.Ho]{Carlos
Hern\'andez--Monteagudo$^{1}$\thanks{E-mail:chm@mpa-garching.mpg.de} and
Shirley Ho$^{2}$\\
$^{1}$Max Planck Institut f\"ur Astrophysik (MPA), Karl Schwarzschild Str.1, Garching bei M\"unchen, D-85741, Germany\\
$^{2}$Lawrence Berkeley National Laboratory, Berkeley, CA 94704, USA
}
\begin{document}

\date{}

\pagerange{\pageref{firstpage}--\pageref{lastpage}} \pubyear{2008}

\maketitle

\label{firstpage}

\begin{abstract}

The peculiar motion of ionized baryons is known to introduce temperature anisotropies in the Cosmic Microwave Background radiation (CMB) by means of the kinetic Sunyaev-Zel'dovich effect (kSZ). In this work, we present an all sky computation of angular power spectrum of the temperature anisotropies introduced by kSZ  momentum of all baryons in the Universe during and after reionization. In an attempt to study the bulk flows of the {\em missing baryons} not yet detected, we address separately the contribution from {\em all} baryons in the Inter Galactic Medium (IGM) and those baryons located in collapsed structures like groups and clusters of galaxies. In the first case, our approach provides a complete, all sky computation of the kSZ in second order of cosmological perturbation theory (also known as the Ostriker-Vishniac effect, OV). Most of the power of OV is generated during reionization, although it has a non-negligible tail at low redshifts, when the bulk of the kSZ peculiar momentum of the halo (cluster + group) population arises.  If gas {\em outside} halos is comoving with clusters as the theory predicts, then the signature of the bulk flows of the {\em missing} baryons should be recovered by a cross-correlation analysis of future CMB data sets with kSZ estimates in clusters of galaxies. For an ACT or SPT type of CMB experiment, all sky kSZ estimates of all clusters above $2\times 10^{14}\; h^{-1}M_{\odot}$ should provide a detection of {\it dark} flows with signal to noise ratio (S/N) of $\sim 10$, (S/N $\sim 2.5-5$ for 2,000 - 10,000 square degrees). Improving kSZ estimates with data from Large Scale Structure surveys should enable a deeper confrontation of the theoretical predictions for bulk flows with observations. The combination of future CMB and optical data should shed light on the dark flows of the nearby, so far undetected, {\em diffuse} baryons.

\end{abstract}


\section{Introduction}

The standard cosmological model predicts the statistical properties and evolution of the large scale 
matter distribution. Initially small inhomogeneities in the matter density field grow under gravity 
in an expanding universe, giving rise to a cosmic web made of filaments, sheets and superclusters of 
galaxies that we see today. Large, comoving flows of matter fall 
onto deep potential wells under gravity, with typical speeds of $\sim 150$ km s$^{-1}$ in the standard 
$\Lambda$CDM cosmological model \citep{wmap08}. These peculiar velocities ( peculiar
with respect to the Hubble flow induced by the universal expansion) cannot be easily measured. 
Indeed, despite recent claims pending for confirmation \citep[e.g.,][]{sashaksz08,watkins08}, 
there is no conclusive observational evidence for such bulk flows in the present universe.

Nevertheless,  the Cosmic Microwave Background (CMB) radiation has provided an accurate measurement 
of these bulk flows when the universe was 380,000 years old ($z\simeq 1,050$). 
Before most of the electrons recombined with the protons and the CMB could propagate freely, 
the electron plasma and the CMB radiation were in thermal equilibrium since Thomson scattering 
kept both fluids coupled. The free electrons Thomson scattered CMB photons, leaving Doppler imprint according to 
their relative velocities caused by the oscillation of the plasma-photon fluid in the gravitational 
potential wells \citep{sunyaev70, husugiyama95}. Since energy and density fluctuations were 
very small at that epoch, the linear theory is able to describe very accurately the anisotropy pattern that was introduced by 
Thompson scattering on the CMB, and this has been confirmed by observations \citep[e.g.][]{BOOMERANG00, TOCO02, VSA03, wmap08}.

At redshift $z\sim 10$ the first stars began to ionize hydrogen in the Intergalactic Medium (IGM), and therefore CMB photons interacted again with the free electrons via Thomson scattering. 
In this scenario, electron peculiar velocities introduced new temperature anisotropies on the CMB. 
Likewise, due to the anisotropic nature of Thomson scattering, the CMB was also linearly polarized 
on the large angular scales, as it has also been confirmed by the WMAP experiment, \citep{wmap08}. 
These epochs describe the last scenario where {\em all} baryons have been observed: at more recent epochs, the amount of baryon matter that we are able to detect is at most of half of what is observed 
during both recombination and reionization \citep{fukpeebles}. This constitutes the {\it missing baryon} problem. 

In this work, and just as in \citet{chmEkSZ}, we investigate bulk flows and
the missing baryons utilizing Thomson scatterings between CMB photons and free
electrons at low redshift.  This mechanism is referred to as the kinetic
Sunyaev-Zel'dovich effect \citep[kSZ, ][]{kSZ}, if it occurs in galaxy
clusters, but it is called the Ostriker Vishniac effect \citep[OV,][]{OV} in
the context of cosmological second order perturbation theory, (for which the
electron density and velocity are continuous fields). We shall follow this
terminology, although in some occasions we may use either when referring to
both effects for simplicity.  We shall restrict ourselves to the impact of
Thomson scattering on the intensity or temperature anisotropies of the CMB,
since the effect on the CMB polarization is already studied in
\citet{chmEkSZ}.  Unlike in \citet{letterdtsz}, we shall not consider the
spectral distortions that the hot fraction of the missing baryons introduce in
the CMB black body spectrum. Previous efforts have been made in the
computation of the OV effect \citep{OV, dodelson95, hu00, castroOV}, some of
them in the context of the non-Gaussian signal they give rise to. In some
recent works, the kSZ in halos has been used as a probe for Dark Energy
\citep{ksky1}, while \citet{moodley_kSZ} have used the gas in
groups to probe the missing baryons, and \citet{lee_kSZ} studies the
prospects of constraining patchy reionization.  Here, we present the first
full sky projection of the temperature anisotropies induced by peculiar
momenta of ionized baryons during and after reionization up to the present: if
applied to the smooth electron distribution our calculations provide a full
description of the OV effect, while when applied to the halo distribution they
describe the contribution of galaxy group/cluster correlation on the kSZ
effect. The formalism also permits the computation of the cross correlation
between the OV and the kSZ effects. At the same time, our computations provide
the linear approximation under which the electron density is approximated by
its average value. We can then compare our code with standard linear theory
Boltzmann codes such as CMBFAST \citep{cmbfast} or CAMB \citep{camb}, as well
as normalize our predictions to the linear fluctuation amplitudes measured by
WMAP. Finally, we analyze our results in the context of future CMB and Large
Scale Structure observations, and provide predictions for the detectability of
bulk flows and missing baryons in the light of those future data sets. The
paper is structured as follows: in Section \ref{sec:2ndpt} we describe the
line of sight projection of the baryon peculiar momentum, and this is
discussed in Section \ref{sec:halo} in the context of its application to the
halo population.  The amplitude of the resulting power spectra are given in
Section {\ref{sec:pspectra}}, and their interpretation in terms of bulk
flow/missing baryon detection is provided in Section \ref{sec:pros}. We
discuss our main results in Section \ref{sec:disc} and conclude in Section
\ref{sec:con}.

\section{The second order kSZ effect (OV effect)}
\label{sec:2ndpt}
The computation of the OV/kSZ effect requires the projection along the line of sight the product of electron peculiar velocity and density. If we consider the spatial variations of the electron density, then this product becomes a second order quantity in cosmological perturbation theory provided that peculiar velocities are very small ($v/c \ll 1$). I.e., the CMB temperature anisotropies introduced by the peculiar motion of the scatterers reads
\begin{equation}
\frac{\delta T}{T_0} (\n) = \int \;d\eta\; e^{-\tau_T}\; a(\eta )\sigma_T \bar{n}_e \left(1+\delta_e\right)   \left( -\frac{\vv_e\cdot\n}{c} \right),
\label{eq:kSZ0}
\end{equation}
with $\tau_T$ the Thomson optical depth 
\begin{equation}
\tau_T = \int \; d\eta\; a(\eta) \sigma_T n_e(\eta ),
\label{eq:tauT}
\end{equation}
$a(\eta ) = 1/(1+z(\eta))$ the cosmological scale factor, $\sigma_T$ the Thomson cross section and $n_e(\eta ) = \bar{n}_e (1+\delta_e)$ is the electron number density in terms of its average value ($\bar{n}_e$) and its spatial fluctuations ($\delta_e$). The symbol $\eta$ denotes conformal time or, equivalently, 
comoving distance.  In \citet{kSZchm} (hereafter HVJS) a full sky 
description of the kSZ generated by the low redshift galaxy cluster population was provided.
However, as noted later by \citet{gordonkSZ}, those computations observed only the diagonal part of the Fourier peculiar velocity tensor $\langle v^i_{\vk} v^{j\;*}_{\vq} \rangle$ (where $\vk$ and $\vq$ are Fourier wave vectors and $i$ and $j$ refer to one of the three spatial components of the peculiar velocity vectors), neglecting all non-diagonal components.  
In Appendix A, we present a full sky computation considering all terms in this tensor, and find that the complete treatment introduces small changes to the results given in HVJS. The results of the Appendix are summarized here.

In brief, the second order statistic for the electron momentum ($p_e \propto n_e v_e$) is proportional to the quantity $\langle n_{e,1}v_{e,1}n_{e,2}v_{e,2}\rangle$. If $n_{e,i}$ is constant, then it simplifies to ${\bar n}_e^2 \langle v_{e,1} v_{e,2}\rangle$, and this is the $vv$ term. 
Furthermore, it is possible that $n_e$ is driven by Poissonian statistics, in which case $\langle n_{e,1}v_{e,1}n_{e,2}v_{e,2}\rangle \propto n_e \langle v_{e,1} v_{e,2}\rangle$ (Poisson term). If however $n_{e,i}$ is correlated to the velocity field, then one must use the cumulant
expansion theorem, and three new terms arise, of which one is positive, another one is zero and the third one is negative. In all cases, the correlation function $C(\n_1,\n_2)$ for each term can be written as Fourier integrals over some wave vector $\vk$ of functions with an angular dependence of either the type $(\kh\cdot \n_1)(\kh\cdot \n_2)$ or $(\n_1 \cdot \n_2)$. Those contributions with $(\kh\cdot \n_1)(\kh\cdot \n_2)$ type dependence will give negligible contribution at the small scales,
i.e., the corresponding angular power spectrum multipoles $C_l$ will go to zero as the multipole $l$ grows. In this limit, only the contributions with the $(\n_1 \cdot \n_2)$ type angular dependence will survive. Let us consider the above more carefully in case by case basis. 

\subsection{The linear terms}

If the electron density is constant,
then kSZ/OV effect is linear and the corresponding power spectrum will contribute to the total with a term that we shall name as {\it vv}. The amplitude of the {\it vv} term is given by
\[
 C_l^{vv} = \left( \frac{2}{\pi} \right) 
\int dk\;k^2 \; \frac{P_m(k)}{k^2} \;\times
\]
\[
\biggl( \frac{(l+1)^2}{(2l+1)^2} {\cal I}^{vv}_{l+1,1}{\cal I}^{vv}_{l+1,2} - \frac{(l+1)l}{(2l+1)^2} {\cal I}^{vv}_{l+1,1}{\cal I}^{vv}_{l-1,2} 
\]
\begin{equation}
\phantom{xxxxxx}
- \frac{(l+1)l}{(2l+1)^2} {\cal I}^{vv}_{l-1,1}{\cal I}^{vv}_{l+1,2}  + \frac{l^2}{(2l+1)^2} {\cal I}^{vv}_{l-1,2}{\cal I}^{vv}_{l-1,2}
\biggr).
\label{eq:cvvtxt}
\end{equation}
In this equation, the terms ${\cal I}_{l,i}^{vv}$ correspond to the projection on multipole $l$ of the line of sight integrals of the redshift 
dependent source term that includes the growth factor of the peculiar velocities ${\cal D}_v(z)$ and the average Thomson visibility function
$\dot{\tau_T} \exp{(-\tau_T)}$ (with $\dot{\tau_T} = \sigma_T \bar{n}_e \;a(z)$ the Thomson opacity, dependent on the Thomson cross section, 
the average electron density ($n_e$) and the scale factor $a(z)$). The label $i$ refers to different lines of sight $\n_i$, $i=1,2$. The linear peculiar velocity growth factor ${\cal D}_v$ can be expressed in terms of the linear density growth factor (${\cal D}_{\delta}$),
\begin{equation}
{\cal D}_v \equiv H(z) \biggl| \frac{d{\cal D}_{\delta}}{dz} \biggr|,
\label{eq:dv_def1}
\end{equation}
and ${\cal D}_{\delta}(z)$ is expressed in terms of the integral
\begin{equation}
{\cal D}_{\delta}(z) \propto  H(z)\int_{z}^{\infty}\frac{dz}{a^2(z) H(z)},
\label{eq:dd_def1}
\end{equation}
with $H(z)$ the Hubble function and ${\cal D}_{\delta}$ normalized to ${\cal D}_{\delta}(z=0) = 1$ .The spatial clustering properties of the peculiar velocity field are encoded in the integral of the velocity power spectrum, 
which is proportional to $P(k)/k^2$ ($P(k)$ is linear matter power spectrum).  Provided that all lines of sight are statistically 
equivalent and for high $l$ there exists an asymptotic value for the spherical Bessel functions $j_l(x)$,
 it is easy to find that at the small scales the $vv$ term vanishes, i.e., 
\begin{equation} 
\lim_{l\rightarrow \infty} C_l^{vv} = 0.
\label{eq:limitCvvtxt}
\end{equation}
This term is linear, and is therefore present in standard Boltzmann codes like CMBFAST or CAMB.

In the case of halo induced kSZ, the discrete nature of halos introduces a different type of anisotropy. 
The actual amplitude of the kSZ is proportional to the number of halos present along the line of sight, 
therefore kSZ fluctuations obey Poisson statistics in this case. The variance of the number of objects is
proportional to its mean, and so the angular power spectrum is also linear in the average halo number density. 
The Poisson term has two contributions, $C_{l,A}^{P}$ and $C_{l,B}^{P}$. The first one is
\[
C_{l,A}^{P} =
\left( \frac{2}{\pi}\right) \int k^2dk\;{\cal J}_1(k) \; \times
\] 
\begin{equation} 
\phantom{xxxxxxxx}
 \biggl( \frac{l+1}{2l+1}{\cal I}^P_{l+1,1} {\cal I}^P_{l+1,2}  +  \frac{l}{2l+1}{\cal I}^P_{l-1,1} {\cal I}^P_{l-1,2}\biggr),
\label{eq:clpA}
\end{equation}
and it does not vanish on the small scales when both lines of sight are statistically equivalent,
although it will drop to zero in the angular range that goes well beyond the typical halo profile (halos are assumed to have no substructure,

The function ${\cal J}_1(k)$ is an angular integral involving the window functions from the Fourier transform of the halo profile 
and the matter power spectrum. The terms ${\cal I}_{l,i}^{P}$ express the multipole integral along the line of sight the visibility function and the velocity growth. They are proportional to $\sqrt{n_h}$ ($n_h$ is the halo number density),
so that their product is linear in the halo density, as Poisson statistics requires.  
On the other hand, the term $C_{l,B}^{P}$ reads
\[
C_{l,B}^{P} =
\left( \frac{2}{\pi}\right) \int k^2dk\;{\cal J}_2(k)\times
\]
\[
\phantom{x}
 \biggl(    \frac{(l+1)^2}{(2l+1)^2} {\cal I}^{P}_{l+1,1}{\cal I}^{P}_{l+1,2} - \frac{(l+1)l}{(2l+1)^2} {\cal I}^{P}_{l+1,1}{\cal I}^{P}_{l-1,2} 
 \]
 \begin{equation}
 \phantom{xxxxx}
- \frac{(l+1)l}{(2l+1)^2} {\cal I}^{P}_{l-1,1}{\cal I}^{P}_{l+1,2}  + \frac{l^2}{(2l+1)^2} {\cal I}^{P}_{l-1,2}{\cal I}^{P}_{l-1,2}  \biggr).
\label{eq:clpB}
\end{equation}
This term vanishes again at high $l$-s,
\begin{equation}
\lim_{l\rightarrow \infty} C_{l,B}^{P} = 0,
\label{eq:limClpa}
\end{equation}
so in this multipole regime $C_l^P \rightarrow C_{l,A}^{P}$.  
The function ${\cal J}_2(k)$ is another $k$ dependent function similar to ${\cal J}_1(k)$.  

\subsection{The second order terms}

Both the $vv$ and the Poisson terms above are linear, and hence their angular power spectra are proportional to the 
linear scalar power spectrum (although we prefer to write them in terms of the linear matter power spectrum). 
Second order terms involve the product of density and peculiar velocity, and thus resulting angular power spectra contain 
terms that can be interpreted as convolutions of the initial scalar power spectrum. In the Appendix,
we demonstrate that three different terms of this nature arise, namely the $v_1d_1-v_2d_2$ term,  the $v_1v_2-d_1d_2$ term and 
the $v_1d_2-v_2d_1$ term. (The integer subscript here denotes one of the two lines of sight). 
The first one turns out to give zero contribution, so for simplicity we shall refer to the latter two as the $vv-dd$ and $vd-vd$ terms, respectively. 

The $vv-dd$ term comes from two contributions,
\begin{equation}
C_l^{vv-dd} = C_{l,A}^{vv-dd} + C_{l,B}^{vv-dd}.
\label{eq:clddvvdecomp}
\end{equation}
The $C_{l,A}^{vv-dd}$ term reads
\[
 C_{l,A}^{vv-dd} = 
 \left(\frac{2}{\pi}\right)\int k^2dk\;  {\cal A}(k) \;\times
 \]
 \[
 \biggl[ {\cal I}_{l+1,1}^{vv-dd} \biggl( \frac{(l+1)^2}{(2l+1)^2}{\cal I}_{l+1,2}^{vv-dd} - \frac{l(l+1)}{(2l+1)^2}{\cal I}_{l-1,2}^{vv-dd}\biggr) + 
 \]
 \begin{equation}
  \phantom{xxxxx}
 {\cal I}_{l-1,1}^{vv-dd} \biggl( - \frac{l(l+1)}{(2l+1)^2} {\cal I}_{l-1,2}^{vv.dd} +  \frac{l^2}{(2l+1)^2} {\cal I}_{l-1,2}^{vv-dd}\biggr)
\biggr],
 \label{eq:clddvva}
 \end{equation}
and, as the $vv$ term above, it vanishes on the small scales. The expression for $C_{l,B}^{vv-dd} $ is
\[
 C_{l,B}^{vv-dd} = 
\left(  \frac{2}{\pi}\right) \int k^2dk\; {\cal B}(k) \; \times
\]
\begin{equation}
  \phantom{xxxxx}
\biggl[ {\cal I}_{l+1,1}^{vv-dd} 
{\cal I}_{l+1,2}^{vv-dd} \frac{l+1}{2l+1} +  {\cal I}_{l-1,1}^{vv-dd} 
{\cal I}_{l-1,2}^{vv-dd} \frac{l}{2l+1} 
\biggr] ,
 \label{eq:clddvvv}
 \end{equation}
and dominates the contribution to $C_l^{vv-dd}$ at high multipoles. Contrary to ${\cal J}_1(k), {\cal J}_2(k)$ above, 
the functions ${\cal A}(k)$ and ${\cal B}(k)$ are angular integrals involving the convolution of two matter power spectra, 
apart from the window functions from Fourier transform of the halo profile.
The terms ${\cal I}_{l,i}^{vv-dd}$ provide the line of sight integral on multipole $l$ of the visibility function, 
the matter bias factor (matter bias factor is set to unity unless when we are considering halos) and the growth factors of 
density and velocities, respectively.

The term $C_l^{vd-vd} $ is very similar to $C_l^{vv-dd} $: it is built upon two different contributions,
\begin{equation}
C_l^{vd-vd} = C_{l,A}^{vd-vd} + C_{l,B}^{vd-vd},
\label{eq:cldvdvdecomp}
\end{equation}
and the first one also vanishes on the small scales, as can be see from this expression when we go to large $l$:
\[
 C_{l,A}^{vd-vd} = 
\left( \frac{2}{\pi}\right) \int k^2dk\; {\cal C}(k) \times
 \]
 \[
  \phantom{xx}
 \biggl[ {\cal I}_{l+1,1}^{vd-vd} \biggl( \frac{(l+1)^2}{(2l+1)^2}{\cal I}_{l+1,2}^{vd-vd} - \frac{l(l+1)}{(2l+1)^2}{\cal I}_{l-1,2}^{vd-vd}\biggr) + 
\]
\begin{equation} 
  \phantom{xxxxx}
 {\cal I}_{l-1,1}^{vd-vd} \biggl( - \frac{l(l+1)}{(2l+1)^2} {\cal I}_{l-1,2}^{vd-vd} +  \frac{l^2}{(2l+1)^2} {\cal I}_{l-1,2}^{vd-vd}\biggr)
\biggr].
 \label{eq:clvdvda}
 \end{equation}
 The second contribution is given by
 \[
 C_{l,B}^{vd-vd} = 
 \left(\frac{2}{\pi}\right) \int k^2dk\; (-{\cal D}(k)) \;\times
 \]
 \begin{equation} 
 \biggl[ {\cal I}_{l+1,1}^{vd-vd} 
{\cal I}_{l+1,2}^{vd-vd} \frac{l+1}{2l+1} +  {\cal I}_{l-1,1}^{vd-vd} 
{\cal I}_{l-1,2}^{vd-vd} \frac{l}{2l+1} 
\biggr] .
 \label{eq:clddvvv}
 \end{equation}
The functions ${\cal C}(k)$ and ${\cal D}(k)$ are again angular integrals which contain the convolution of two power spectra, 
spherical Bessel functions and window functions associated with the size of the regions where peculiar velocities are measured.

\section{Gas in halos versus gas in the IGM}
\label{sec:halo}

\begin{figure}
\begin{center}
        \epsfxsize=7cm \epsfbox{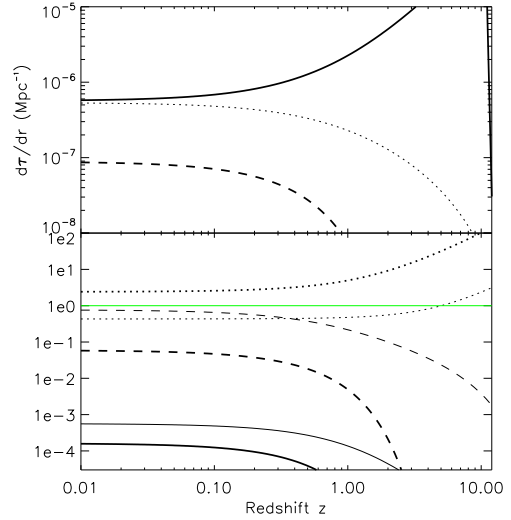}
\caption[fig:taudots]{{\it Top:}  Opacities in a WMAP V Universe due to a smooth distribution of electrons (solid line), 
all halos more massive than $5\times 10^{5} h^{-1}M_{\odot}$ (dotted line) and all halos more massive than $5\times 10^{13} h^{-1}M_{\odot}$ (dashed line). {\it Bottom:}  Thin lines refer to the $M>5\times 10^{5} h^{-1}M_{\odot}$ halo population, thick lines to $M>5\times 10^{13} h^{-1}M_{\odot}$. {\it Dotted lines:} Bias factor evolution versus redshift. {\it Solid lines:} Volume fraction filled by each halo population. {\it Dashed lines:} Product of volume fraction times electron overdensity (with respect to background).  
   }
\label{fig:taudots}
\end{center}
\end{figure}

Before we show the amplitudes of the OV/kSZ power spectra presented in the previous section, we compare 
the effect of Thomson scattering as it arises in collapsed halos to the OV generated in the intergalactic medium (IGM).
It is relevant to understand under which conditions would collapsed gas in halos contribute more in terms of CMB angular anisotropies than the 
gas in the IGM.
In both scenarios, Thomson scattering introduces anisotropies according to
\begin{equation}
\frac{\delta T}{T_0} (\n ) = \int d\eta \; \dot{\tau_T} \left( -\frac{\vv\cdot\n}{c} \right),
\label{eq:kSZ}
\end{equation}
where we have approximated the Thomson visibility function $\Lambda = \dot{\tau_T} \exp{(-\tau_T)}$ by the opacity,
 $\Lambda \simeq \dot{\tau_T}$, since $\tau_T$ is $\sim 0.09$ in the redshift regime of interest \citep{wmap08}.

We assume that the peculiar velocity of gas with respect to the CMB is very similar for both gas in IGM and halos, the 
comparison of the OV and kSZ effects must then be focused on the optical depths, i.e., in the line of sight 
integral of $\dot{\tau_T} $ in each case. As shown in HVJS, the effective opacity generated in halos at a given epoch is given by 
\[
\langle \dot{\tau}_T\rangle_{halo} (z) = \int dM \; \frac{dn}{dM} \dot{\tau}_{T,halo} (M,z) V(M,z) = 
\]
\begin{equation}
\phantom{xxxxxxx}
\int dM \; {\dot \tau}_{T,halo} (M,z) f_{vol}(M,z),
\label{eq:tauhalos}
\end{equation}
where $dn/dM$ is the halo mass function (we choose Sheth and Tormen \citep[ST,][]{ST}, $V(M,z)$ is the proper volume occupied by gas 
in those halos, and $\dot{\tau}_{T,halo}(M,z)$ is the opacity generated by a halo of mass $M$ at redshift $z$. 
The effect is in the end sensitive to the total number of electrons 
in a given halo population, 
and can be expressed by 
the average halo optical depth times the volume fraction of halos.

The halo population opacity can be compared directly to the opacity associated to the smooth distribution of {\em all} electrons.
The top panel of Figure (\ref{fig:taudots}) displays the opacity in three scenarios: {\it (i)} a smooth distribution of electrons (solid line)\footnote{We assume a sudden reionization model
within WMAP V cosmology, for which $\tau_T = 0.084\pm 0.016$ and $z_{reio}=10.8\pm 1.4$}, {\it (ii)} electrons within all halos of mass 
greater than $5\times 10^{13} h^{-1}M_{\odot}$ (dashed line), and {\it (iii)} electrons within all halos of mass greater than 
$5\times 10^{5} h^{-1}M_{\odot}$ (dotted line). The most massive halo population provides a very low opacity due to the small 
amount of cosmological volume it takes. The latter halo population contains almost all the electrons in the Universe, and therefore 
its opacity is very close to the one in case {\it (i)}\footnote{According to the halo model, {\em all} mass in the Universe must be 
found in collapsed halos}.

We have ignored the effect of the bias until now. Dark matter halos can either be biased or anti-biased tracers of the underlying 
matter distribution, and we need to include the bias factor when we compute the angular anisotropies. 
In the bottom panel of Figure (\ref{fig:taudots}) the bias factors for the two halo populations considered ($M  > 5\times 10^{5} h^{-1}M_{\odot}$ and $M > 5\times 10^{13} h^{-1}M_{\odot}$) are shown by the dotted thin and dotted thick lines, respectively. These bias factors were computed by using the expressions of \citet{mowhite96}, and integrating them with the ST mass function:
\begin{equation}
b(z) \equiv \int dM\frac{dn}{dM}\; b(M,z) \bigg/ \int dM\frac{dn}{dM}.
\label{eq:bvsz}
\end{equation}
Although at high redshifts bias factors grow above unity, the volume fraction occupied by the halo population are very low: 
this is shown by bottom solid lines, (thin line for $M  > 5\times 10^{5} h^{-1}M_{\odot}$, 
thick line for $M > 5\times 10^{13} h^{-1}M_{\odot}$). The volume fraction is so low that it cannot be compensated by 
gas overdensity within halos: the product of gas overdensity and the volume fraction is below unity at all redshifts (as shown by the thin 
($M > 5\times 10^{5} h^{-1}M_{\odot}$) and thick ($M > 5\times 10^{13} h^{-1}M_{\odot}$) dashed lines). 
The thin solid green line shows the level of unity. 

According to these results, it seems that the anisotropy due to halos will always be below the level of anisotropy induced by 
a smooth and continuous distribution of electrons. There are however two caveats: 
{\it (i)} Halos might show bias in their peculiar velocities, although this is unlikely to change the picture by 
more than a 30\%-40\% \citep{peel06}, {\it (ii)} on the small scales, halos follow Poissonian statistics, and this 
will add a considerable amount of anisotropy, as we shall see below. 
These results differ from previous works \citep[e.g.,][]{hu00, castroOV} where the power spectrum for kSZ 
in non linear structures had been computed. In previous works, the matter power spectrum was simply 
replaced by the non linear power spectrum, but no changes were introduced in the related opacity. 
No attention was paid to the Poisson term either, which, in the case of halos, turns out to be dominant. 
This difference must be motivated by the fact that, they were computing the combined effect of both linear and non-linear 
density perturbations at high redshift. Here we try to separate the OV contribution from the one generated in halos, 
and the contribution to the latter arises mostly at low ($z<2$) redshift.

\section{The Power Spectrum of the OV/kSZ effect}
\label{sec:pspectra}

\subsection{Normalization to WMAP observations}
The WMAP satellite has provided all sky measurements of the temperature and polarization anisotropies of the CMB \citep{wmap08}. 
This fluctuation field is so small that should provide a normalization for the linear level of departure from perfect anisotropy 
(higher order contributions should be negligible). 
On the large scales, the theory predicts that the CMB temperature anisotropies must caused mainly by two physical phenomena: 
gravitational redshift/blueshift of CMB photons due to the passage through different (and eventually evolving) gravitational
 potentials, and Thomson scattering induced by free electrons during and after reionization.
This theory predicts the relative weight of these two contributions, and WMAP measurements constitute 
accurate measurements of the {\em combined effect} of these two contributions. Therefore,
if the theory is correct, by using WMAP observations it is possible to estimate the amplitude of each of 
the two contributions dominating at the large angular scales. In particular, WMAP data can be used to normalize
the level of anisotropies generated by Thomson scattering during and after reionization. 

\begin{figure}
\begin{center}
        \epsfxsize=7cm \epsfbox{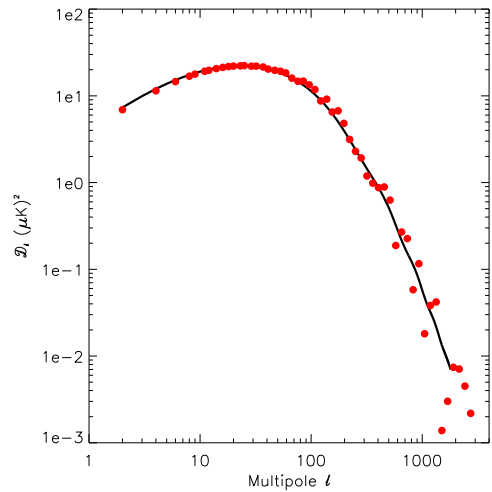}
\caption[fig:norm]{Comparison of the WMAPV-normalized prediction of the $vv$ term by CMBFAST (solid line) with our computation (red filled circles).
   }
\label{fig:norm}
\end{center}
\end{figure}

In Figure (\ref{fig:norm}), the black solid line displays the $vv$ term which accounts for the Thomson scattering contribution to the low $l$ CMB anisotropy, (see Equation (\ref{eq:cvvtxt})). This has been produced by feeding the cosmological  parameter set given for WMAPV+SNALL+BAO \footnote{See  {\tt http://lambda.gsfc.nasa.gov/product/map/dr3/params}} to a modified version of CMBFAST. This code provides both the total amplitude of the angular power spectrum and the $C_l^{vv}$ term. 
The total power spectrum was normalized to fit WMAPV observations, and the $vv$ term was scaled accordingly. As shown by the solid line in  Figure (\ref{fig:norm}), the maximum of the $vv$ angular band power spectrum is reached at $l\sim 20-30$, with an amplitude close to $\sim 22$ $(\mu K)^2$. Note that here power spectrum amplitudes are given in terms of quantity ${\cal D}_l \equiv l(l+1)C_l/(2\pi)$. The filled red circles display the result of our computation of Equation (\ref{eq:cvvtxt}): the accurate computation of this term is particularly difficult, since at large $l$'s it drops to zero as the vanishing difference of several integrals. Small numerical errors in the computation of these integrals introduce some scatter with respect to the CMBFAST prediction, but only at moderate and high $l$'s. For $l<100$, the accuracy of our code is better than 8\%, and we should expect a similar or better level of accuracy for all other terms which do not vanish as the $vv$ term, (we remark that all other terms become dominated at high $l$'s by at least a non-vanishing contribution). Note that this comparison allows us normalize the output of our code not only for the linear predictions,but also for the second order predictions of $vv-dd$ and $vd-vd$. Therefore, with the exception of the percent level uncertainty of our integrals, all sources of uncertainties are associated to the modeling of the mass function, the bias factor and the halo profile description.

\subsection{Results}

\begin{figure*}
\centering
\plotancho{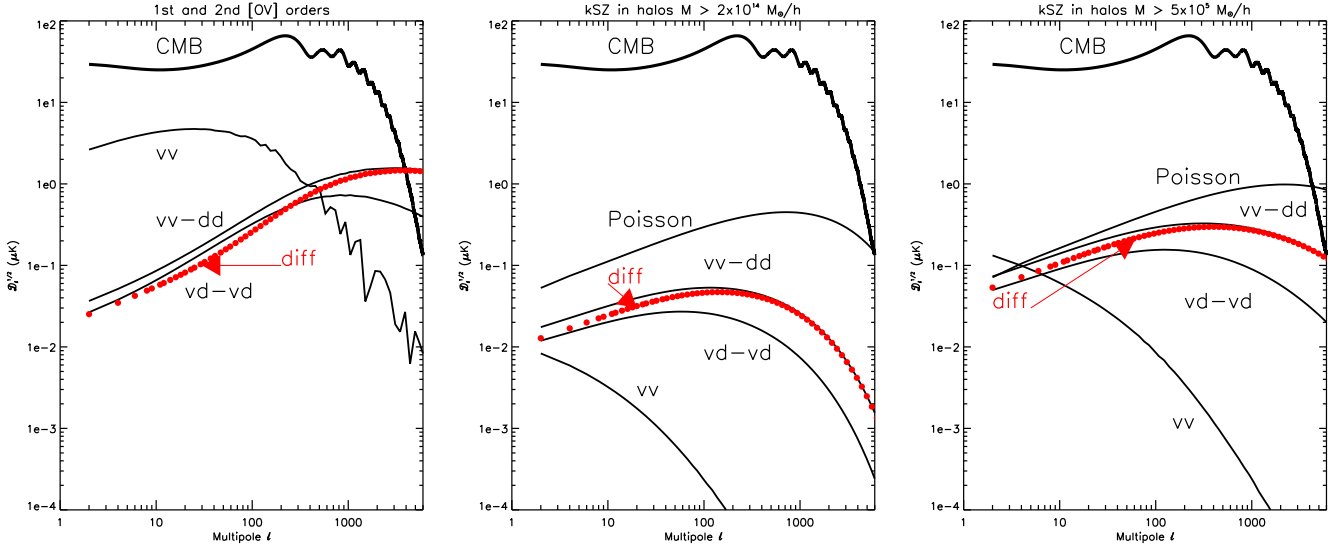}
\caption[fig:cls]{
Angular power spectra of the OV/kSZ effect in $\mu K$ units, (${\cal D}^{1/2}_l \equiv \sqrt{l(l+1)C_l/(2\pi)}$). The red filled circles provide the difference in amplitudes between the $vv-dd$ and the $vd-vd$ term, since the latter is negative.  {\it Left panel:} Anisotropy amplitude generated by a smooth distribution of electrons during and after reionization (OV effect). {\it Middle panel:} First and second order kSZ amplitude generated a cluster population with $M> 2\times 10^{14} \;h^{-1}M_{\odot}$.
 {\it Right panel:} First and second order kSZ amplitude generated by a cluster population with $M> 5\times 10^{5} \;h^{-1}M_{\odot}$.
 }
\label{fig:cls}
\end{figure*}

The amplitude of the OV/kSZ power spectra in different scenarios are displayed in Figure (\ref{fig:cls}).
Note that ${\cal D}_l^{1/2} = \sqrt{l(l+1)C_l/(2\pi)}$ in $\mu K$. The left panel considers all electrons in 
a diffuse and continuous phase, and should be a fair description of the electron distribution during reionization. This case adopts the opacity displayed by the thick solid line in the top panel of Figure (\ref{fig:taudots}), for which the anisotropy is mostly generated at $z\sim 6-10$. The linear $vv$ term we use to calibrate the amplitude of the fluctuations peaks at $\sim 4-5$ $\mu K$ for $l\sim 10 - 20$, but then drops rapidly for $l>100$. On the contrary, the sum of the $vv-dd$ and $vd-vd$ terms (which corresponds to what is commonly understood as the OV effect) becomes dominant at $l\sim 200$, and shows a rather flat pattern on the small scales. The OV term crosses the linear power spectrum generated during recombination at $l\sim 4,000$, with an amplitude between $1-2$ $\mu K$. 
Provided that there is no way to spectrally distinguish the power due to OV during reionization from the power originated at the last scattering surface, the crossover of the OV over the linear power spectrum is particularly significant, since it provides a direct observable of the reionization epoch. Its amplitude is within the nominal sensitivity of third generation CMB experiments like ACT \citep{ACT} or SPT \citep{SPT}.

Indeed, the amplitude of the OV effect is considerably higher than the amplitude corresponding to the $vv-dd$ and $vd-vd$ terms generated by halos. It is well above the anisotropy level generated by the galaxy cluster population which experiments like ACT and SPT are targeting ($M> 2\times 10^{14} \;h^{-1}M_{\odot}$, middle panel in Figure (\ref{fig:cls})). Even for $M>5\times 10^5 \; h^{-1}M_{\odot} $, the second order kSZ terms lie about one order of magnitude below the OV contribution
 at $l\sim 3,000$ (which translates into two order of magnitudes in the power spectrum).
This proves that most of the power of the OV is actually generated at high redshifts, since the OV contribution from low redshifts is well sampled by the kSZ corresponding to masses $M>5\times 10^5 \; h^{-1}M_{\odot} $ (note that the opacities of the OV and the kSZ for this halo population are very similar at low $z$, say, at $z\sim <1$, see Figure (\ref{fig:taudots})). The difference in the amplitudes of the OV and halo kSZ contributions can therefore be assigned to $z>1$.

However, by looking at the right panel of Figure (\ref{fig:cls}) it becomes clear that in order to distinguish the high-z OV contribution from the low-z kSZ halo contribution one must take into account the Poisson term.This can be achieved by masking out all massive, clearly detected halos: after removing the Poisson contribution from clusters above $2\times 10^{14} \;h^{-1}M_{\odot}$, the Poisson term shown in the right panel of Figure (\ref{fig:cls}) drop by a $\sim 40$\%, which, according to our computations, should be below the OV level. We must note that the halo peculiar velocities assumed here are close to the linear limit (we have introduced only a mild bias of 30\%),but reality might be very different due to the presence of thermal, non linear velocities which would boost the kSZ amplitudes in halos. However, to this point, there is very little observational evidence for these non-linear component (as well as for the linear one). 

\section{Prospects for detecting bulk flows/missing baryons in future CMB data}
\label{sec:pros}
\begin{figure*}
\centering
\plotancho{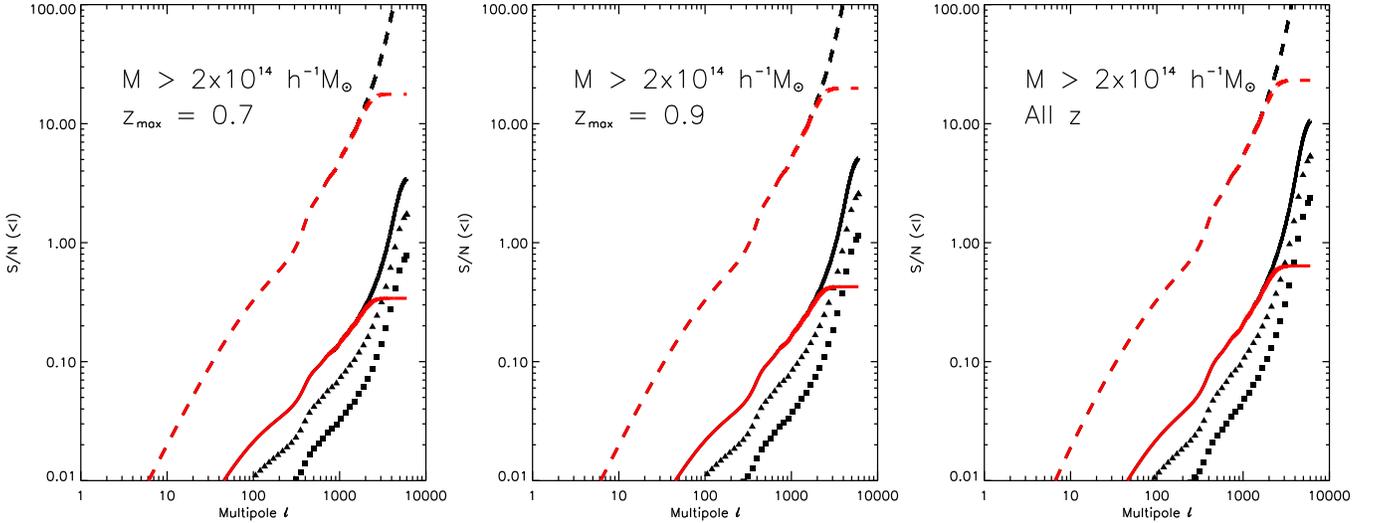}
\caption[fig:s2n]{S/N achieved by cross correlating a kSZ/peculiar velocity template with an all sky CMB map. The kSZ template is built with estimates of peculiar motions in all clusters more massive than $2\times 10^{14} h^{-1} M_{\odot}$. Dashed line refers to the case in which cluster are included in the cross-correlation analysis, whereas solid lines display the case where known halos are masked on the CMB maps, so that only gas outside clusters contributes to the cross-correlation. Black line refers to an ACT-like experiment, red line refers to an experiment with the sensitivity and angular resolution of Planck's 217 GHz channel. Filled triangles (squares) refer to an ACT-like experiment covering 10,000 (2,000) square degrees. The three panels refer to different survey depths: 
$z_{max} = 0.7$ (left), $z_{max} = 0.9$ (center), and all $z$ (right).
}
\label{fig:s2n}
\end{figure*}

Of all kSZ terms shown in Figure (\ref{fig:cls}), one of the easiest to observe is the Poisson term at $l\sim 2,000 - 3,000$. 
Its amplitude is almost as high as the combination of the $vv-dd$ and $vd-vd$ terms from the OV and more importantly, 
it is associated with massive halos that can be identified at other frequencies. Nevertheless, one must note that measuring 
the kSZ Poisson term at positions of galaxy clusters does not properly test the theoretical predictions of the model on the bulk flows. 
This term is sensitive to the average velocity dispersion of those objects (which is likely to be contaminated by non-linear
 contributions) but tells nothing about the spatial correlations of the peculiar velocity field. 

In the context of the standard model, it was shown in HVJS that typical correlation lengths for bulk flows are about 20 -- 40 $h^{-1}$Mpc 
in comoving units. This means that all electrons within a sphere of radius from $20$-$40$ $h^{-1}$Mpc share a similar peculiar velocity. 
If this is an accurate description of reality, then one can use the kSZ detected in the most massive halos as a template for the 
peculiar velocity field in their surroundings.  This idea was used in \citet{chmEkSZ} to track the missing baryons by 
kSZ -- E polarization mode cross-correlation, and can also be applied in this context. 
Let us assume that future high resolution CMB experiments provide accurate measurements of the kSZ in halos more massive than, say,
 $2\times 10^{14} \; h^{-1}M_{\odot}$, so that a template of the kSZ on the sky (hereafter $\vm_{kSZ}$) can be built.
 If this template is then cross-correlated with the CMB sky (hereafter $\vt_{CMB}$), then not only does signal come from the known clusters, 
but also from the surrounding gas. Fundamental sources of noise for this cross-correlation are, to first order, the primordial CMB fluctuations 
and the instrumental noise (which usually becomes relevant on the scales of the PSF).
 This cross-correlation can be picked-up via a matched filter (MF) approach,
\begin{equation}
\alpha = \frac{\vt_{CMB}^t\vC^{-1}\vm_{kSZ}}{\vm_{kSZ}^t\vC^{-1}\vm_{kSZ}},
\label{eq:mf1}
\end{equation}
where the CMB signal is assumed to intrinsically contain some kSZ component proportional to the kSZ template,
i.e., $\alpha \vm_{kSZ}$. The covariance matrix $\vC$ contains the noise sources (namely primordial 
CMB and instrumental noise). The signal to noise ratio (S/N) of this cross correlation is given by
\begin{equation}
\frac{S}{N} = \frac{\vt_{CMB}^t\vC^{-1}\vm_{kSZ}}{\sqrt{\vm_{kSZ}^t\vC^{-1}\vm_{kSZ}}}.
\label{eq:s2n}
\end{equation}
If the kSZ template is written in multipole space, then the last equation on the S/N can be written 
for each multipole $l$. In this basis,  the covariance matrix $\vC$ is diagonal for white instrumental 
noise $N_l$, $(\vC)_{l,l'} = (C_l^{CMB} + N_l) \delta_{l,l'}$, and Equation (\ref{eq:s2n}) reads
\begin{equation}
\left(\frac{S}{N}\right)_l = \frac{ (2l+1) \left( C_{l,kSZ}^{h,h}+C_{l,kSZ}^{h,OV}\right)\left(C_l^{CMB} + N_l\right)^{-1}} 
{\sqrt{(2l+1)C_{l,kSZ}^{h,h}\left(C_l^{CMB} + N_l\right)^{-1}}}.
\label{eq:s2nvsl}
\end{equation}
For non-full sky coverage ($f_{sky}<1$), there are not $(2l+1)$ degrees of freedom per multipole $l$, but roughly $(2l+1)f_{sky}$. In this case, the last equation should be multiplied by $f_{sky}^{1/2}$. 
The term $C_{l,kSZ}^{h,h}$ accounts for the halo-halo kSZ auto-correlation (i.e., the Poisson, $vv$, $vv-dd$ and $vd-vd$ terms), whereas $C_{l,kSZ}^{h,OV}$ expresses the cross-correlation between the kSZ induced in halos and the OV generated by the continuous electron distribution on the large scales (i.e., mostly outside halos). This cross-power spectrum can be easily computed by introducing, 
in our angular cross-correlation function $C(\n_1,\n_2)$, the halo visibility function in the $los$ integrals ${\cal I}_l$-s along direction $\n_1$, and the OV visibility function along direction $\n_2$, (see Appendix). It is worth to stress that in this case there is no contribution from the Poisson term, since this term appears exclusively for the halo auto correlation. If we are searching for the kSZ signal of the gas {\em outside} halos, one should mask the known clusters, which is equivalent to dropping the $C_{l,kSZ}^{h,h}$ term in the numerator of Equation (\ref{eq:s2nvsl}). 

The result of adding all the S/N below a given multipole $l$ is shown in Figure (\ref{fig:s2n}). Black lines refer to an ACT-SPT like all-sky CMB experiment with a sensitivity level of $\sim 1 \mu K$ per resolution element of $\sim 1$ arcmin. 
The red color lines (both solid and dashed) display results for (all sky, $f_{sky}=1$) Planck's HFI 217
GHz channel. If kSZ clusters are not masked, then the cross-correlation test yields a very high S/N, 
as displayed by the dashed lines. Instead, if known kSZ clusters are masked, then the S/N drops to a level than only an ACT-SPT type CMB experiment would be able to detect it, since it mostly comes from the small angular scales. This is shown by the solid lines, which depict the ideal case of full sky coverage ($f_{sky}=1$). On the contrary, filled triangles and squares refer to an ACT-SPT type experiment covering 2,000 and 10,000 square degrees, respectively. We see that these scenarios lie in the limit of detectability.

The latter case describes the situation in which kSZ measurements in galaxy clusters are used to build the kSZ template and search for the baryons in the IGM. The requirements on the sky coverage, angular resolution and instrumental sensitivity are demanding if a kSZ detection of the missing baryons is to be provided. Since the OV is mostly generated at moderately high redshifts, so if the cluster sample 
does not extend deep enough, then the S/N drops accordingly (different survey depths are considered in the different panels of Figure (\ref{fig:s2n})). Note that we do not expect to find massive clusters at high (say $z>4$) redshifts. 

If kSZ measurement in halos are not available, it has been suggested by \citet{HoSDSS} that kSZ templates may be built from the large scale matter distribution.  Present reconstruction algorithms like e.g., ARGO \citep{ARGO} seem to reconstruct peculiar velocities accurately in scales of interest.
This velocity field reconstruction could be an additional way to obtain the kSZ templates, and to detect the kSZ in clusters. In this case the S/N is much higher (dashed lines), and there is no need to have deep surveys: most of the S/N of the angular cross-correlation is coming from the Poisson term. This term does not depend critically on depth, but simply relies on the discrete nature of halos. 
However, since the kSZ template would be a prediction of the model based upon the large scale structure observations, this exercise would still provide a test for the cosmological theory of bulk flows.

\section{Discussion}
\label{sec:disc}

Our formalism allows us to study consistently the OV/kSZ temperature anisotropies introduced by both the gas collapsed in halos and 
the gas present in the IGM, regardless of the redshift range of relevance in each case. Our computations differ from previous ones in 
mainly two aspects: {\it (i)} ours are full sky calculations, where no flat sky/Limber approximation have been invoked and 
therefore predictions apply to all multipole range, and {\it (ii)} the kSZ computations take into account the fact that halos 
above a given mass threshold host only a fraction of the total mass, and hence the amplitude of the kSZ they give rise to is 
limited. 
Indeed, even after accounting for the enhanced clustering of the massive halo population, we find that the kSZ generated by groups and clusters
is below the OV anisotropies produced during and shortly after reionization. Only the Poisson term induced by the halo population reaches an 
amplitude that is comparable to the OV $vv-dd$ term, although it should be possible to partially remove it by masking out the most massive halos. 
By excising all halos above $2\times 10^{14} h^{-1} M_{\odot}$, the Poisson term should drop by a $\sim$40 \% (in the linear units 
of Figure (\ref{fig:cls})).  Distinguishing these two contributions may be a relevant issue, since the kSZ is mostly generated 
at low redshifts ($z<3-4$), whereas the OV is one of the very few existing probes of the reionization epoch  ($z\in [5,12]$). 
This gives significance to the cross-over of the total OV power (red filled circles in the left panel of Figure \ref{fig:cls}) above 
the intrinsic linear CMB power spectrum at $l\sim 4,000$. We remark that this OV computation is merely second order perturbation theory, 
and since the linear level of perturbations has been normalized via, e.g. WMAP observations, there are no free parameters for our (second order) OV predictions. Nevertheless, our computations do not observe patchy reionization, that is, the extra amount of anisotropy 
introduced by a non fully ionized IGM at $z>10$. \citet{zahnOV} claim that this phase of reionization should introduce 
fluctuations comparable in amplitude to the OV, and this would provide further relevance to the study of the anisotropies
in these angular scales. 

This work is also motivated by the search of the missing baryons via their peculiar motions with respect to the CMB. 
If predictions from linear theory are correct, the baryons should comoving on scales of 20 -- 40 $h^{-1}$Mpc, and 
hence the kSZ pattern generated in clusters and groups of galaxies should be very similar to that generated by the surrounding gas,
 and can be used to probe the latter. However, it is not clear yet to what extent kSZ measurements in future CMB surveys 
like ACT or SPT will be contaminated by the intrinsic CMB and point source emission. In the ideal scenario in which errors
 in the kSZ estimates are unimportant for all clusters more massive than $2\times 10^{14} h^{-1} M_{\odot}$, then it is clear
 that the signature of missing baryons can be unveiled if the linear theory predictions on bulk flows are correct.
One can consider to combine them with peculiar velocity field reconstructions obtained 
from LSS surveys, in which accuracy would be required at relatively large scales (10--20 $h^{-1}$Mpc). 
It should be noted that estimates of the kSZ would be needed for each bulk flow, and this can be achieved 
by combining kSZ/peculiar velocity estimates from different halos belonging to the same filament/supercluster. 

Our predictions have larger uncertainties in the kSZ case than in the OV case. We have assumed a Sheth-Tormen mass
function for the halo abundance versus mass and redshift. We also model their physical size and their gas profile, and 
more importantly, we have adopted the Mo \& White prescription for their clustering bias.
When comparing our results with previous works, we find that our kSZ power spectra are at the level of \citet{bjoernkSZ} both 
in shape and amplitude: when imposing the same mass threshold ($M>5\times 10^{13} h^{-1})M_{\odot}$, our kSZ power spectrum is 
within a factor of 2, despite our different choice for $\sigma_8$, (0.817 versus 0.9). Further, we are introducing 
an effective bias for the halo velocities of 1.3, while it is not clear to us how close are halo peculiar velocities compared to
linear theory expectations in the simulation they used. As \citet{bjoernkSZ} pointed out, there may be a significant amount of
extra kSZ power at smaller scales due to the presence of sub-structure within the halos, \citep{volkerkSZ}. 
This could partially be avoided by smoothing the signal within the halo solid angle, although part of this excess 
could be associated to a thermal (non-linear) component of halo velocities.
The addition of all this extra power could have an impact on the S/N estimations of Figure (\ref{fig:s2n}), 
which increases significantly at high $l$-s. A lower mass for the kSZ estimation would then be required in order to achieve 
the same S/N limit. However, despite all these sources of uncertainty, we conclude that future kSZ  and LSS surveys would
shed some light in the problematic of missing baryons and bulk flows in the late universe.

\section{Conclusions}
\label{sec:con}
We have revisited the problem of the secondary anisotropies in the CMB introduced by the peculiar motion of baryons during and after reionization. We have addressed this problem in the context of the search for the missing baryons and the measurements of bulk flows at low redshifts. We have paid particular attention in distinguishing the different sources of anisotropy, according to their redshift, in an attempt to isolate the kSZ/OV generated by gas in filaments and superclusters at low redshift from its counterpart arising during reionization.
For this purpose, we present the first all sky projection of the peculiar momentum of baryons from reionization to the present epoch. We consider two different scenarios: the OV generated by a smooth distribution of {\em all} electrons, and the kSZ generated by only those baryons located in collapsed structures. 

The former provides the largest amplitude, and contributes mostly during reionization ($z>6$). Since it corresponds to second order in perturbation theory, WMAP observations at the linear level already provide a normalization. Once those observations fix the set of cosmological parameters, the predictions at second order should also be fixed: the OV should cross over the intrinsic linear CMB temperature angular power spectrum at $l\sim4,000$, with an amplitude close to $1.5\; \mu K$ (for a sudden reionization scenario at $z_{reio}=10.8$ and $\tau = 0.084$). The amplitude of the OV should be above the angular anisotropy level induced by the kSZ in the halo population. The discrete character of halos (and the Poisson statistics associated to it) gives rise to angular fluctuations that constitute the main contaminant (in terms of kSZ/OV effects) to the signal generated during reionization. By masking out all halos more massive than $2\times 10^{14} \;h^{-1} M_{\odot}$, this Poisson term should drop to a 
level of $\sim 30 \%$ of the OV amplitude, leaving room for this window to the reionization epoch. 

The terms accounting for the velocity -- velocity and velocity -- density correlation of the halo population are well below the Poisson-induced anisotropy level, but are relevant since those terms express the correlation of velocities of halos {\em with the velocities of the surrounding gas}. We propose using kSZ estimates in halos (to be obtained by ACT or STP-type experiments) in order to build peculiar velocity templates to be cross-correlated to future CMB data. If the gas surrounding halos is co-moving with them as the theory predicts, then a signal should arise from this test {\em even after masking the contribution from the halos on the CMB maps}. An ACT or SPT-like CMB experiment covering 10,000 (2,000) square degrees should achieve a S/N of 5 (2.5). The detection of this cross-correlation would provide evidence for the presence of missing baryons in the late universe {\em and}  confirm the predictions of the theory of cosmological bulk flows. In case there are no quality kSZ estimates at the halo angular positions, then, following \citet{HoSDSS}, it would also be possible to build peculiar velocity templates from future deep LSS surveys. These would provide an estimate of the density field that can be inverted (within our cosmological frame) into a peculiar velocity template. By cross-correlating this template with future CMB maps one should be able to detect the kSZ and bulk flows at high significance, {\em if} our understanding of cosmological bulk flows is correct. In summary, cross-correlation studies of future CMB and kSZ/LSS data becoming a promising tool in the characterization of bulk flows and missing baryons in the low redshift universe.

\section*{Acknowledgments} 
We thank D.N.Spergel for useful discussions. CHM acknowledges fruitful discussions with S.D.White.

\bibliographystyle{mn2e}
\bibliography{mjh}

\appendix
\onecolumn
\section{The 2-point angular OV/kSZ correlation function}

This Appendix outlines the computation of the full sky angular power spectrum of the projected peculiar momentum of baryons. It completes the appendixes B and C of HVJS in the sense that it considers the full peculiar velocity tensor $\langle \vv_{\vk,i} \vv_{\vq,j}^*\rangle$ (and not only its diagonal part, as Equation B1 of HVJS implies).\\

We follow the approach of HVJS for integrating kSZ temperature anisotropies along the line of sight. For a given population of halos, the kSZ introduced along the direction $\n$ reads
\begin{equation}
\frac{\delta T}{T_0}(\n ) = \int dr \; a(r) \sigma_T\sum_j W_{gas}(\vrv-\vx_j) n_{e,j} \left(-\frac{\vv_j\cdot \n}{c}\right) = \int dr \; a(r) \sigma_T \int d\vx \;dM\; W_{gas}(\vrv-\vx) n_{e}(M,r) \left(-\frac{\vv(\vx)\cdot \n}{c}\right)
\frac{dn}{dM}(\vx).
\label{eq:los1}
\end{equation}
The sum over $j$ indicates sum over halos whose central electron density is given by $n_{e,j}$ and gas spatial profile by $W_{gas}(\vx)$. The coordinate $r$ is in comoving units and can be regarded as our look back time coordinate and argument of the scale factor $a(r) \equiv 1/(1+z(r))$. The sum  over $j$ can be converted in an integral by introducing the halo mass function $dn/dM$. For notation simplicity, we shall implicitly integrate halo properties over the halo mass, dropping the mass dependence, and renaming the halo number density as $n_h(\vx)$. This transforms the integral in a convolution of the quantity 
\begin{equation}
\xi (\vx) \equiv n_h(\vx) \; \left(-\frac{\vv(\vx)\cdot\n}{c}\right) \equiv n_h (\vx ) \beta (\vx )
\label{eq:xidef}
\end{equation} 
with the profile $W_{gas}(\vx)$:
\begin{equation}
\frac{\delta T}{T_0}(\n ) = \int dr \; a(r) \sigma_T n_{e,h}(r_1)\left( W_{gas} \star \xi\right) (\vrv ) = \int dr \; a(r) \sigma_T n_{e,h}(r_1) \int \dthreek \; W_{gas,\vk} \xi_{\vk}\; e^{-i\vk\cdot \vrv}.
\label{eq:los2}
\end{equation}
The average central halo electron density is denoted by $n_{e,h}$.
The two point angular correlation function can then easily be written as
\begin{equation}
C(\n_1,\n_2) = \int dr_1 \; a(r_1) \sigma_T n_{e,h}(r_1) \int dr_2 \; a(r_2) \sigma_T n_{e,h}(r_2)\; \dthreekuno \dthreekdos\; e^{-i\vk_1\cdot\vrv_1 + i\vk_2\cdot\vrv_2}\;W_{gas,\vk_1} W_{gas,\vk_2}^* \langle \xi_{\vk_1} \xi_{\vk_2}^*\rangle.
\label{eq:c121}
\end{equation}
The ensemble can be expressed in terms of the Fourier modes of the density and peculiar velocity field. Provided that $\xi(\vx )$ is the product of two quantities in real space, its Fourier counterpart equals
\begin{equation}
\xi_{\vk} = \int \dthreeq \beta_{\vq} n_{h,\vk-\vq}.
\label{eq:xikdef}
\end{equation}
Therefore, the ensemble average ($\langle ... \rangle$) in Equation (\ref{eq:c121}) becomes
\begin{equation}
\langle \xi_{\vk_1} \xi_{\vk_2}^* \rangle = \int \dthreequno \dthreeqdos \; \langle \beta_{\vq_1} n_{h,\vk_1-\vq_1} \beta_{\vq_2}^* n_{h,\vk_2-\vq_2}^* \rangle.
\label{eq:xips}
\end{equation}
One must address separately the different contributions on which this power spectrum can be decomposed, and this is the subject of the rest of the Appendix. \\

In the rest of the Appendix, computations will be referred to a halo population. However, these same computations may be referred to a continuous and smooth distribution of electrons if certain changes are introduced in the equations. These changes involve: {\it (i) } the gas profile functions $W_{gas,\vk}$, the Fourier window functions associated to the peculiar velocity and density of halos ($W_{v,\vk}, W_{\delta,\vk}$, to be introduced below) and the central halo electron densities $n_{e,h}$ must be dropped;  {\it (ii)} the halo number density $n_{h,\vk}$ must be substituted by the  electron number density field $n_{e,\vk}$; and {\it (iii)} the halo clustering and velocity biases ($b, b_v$, to be introduced below) must be set to unity. This continuous description of the electron field should be accurate when estimating the projection of the peculiar momentum of {\em all} baryons: the amount of collapsed baryons at high redshift is a very small fraction of the total, and even at present, less than a quarter of the total amount of baryons are found in collapsed halos more massive than $2\times 10^{14}$ $h^{-1}M_{\odot}$. Regardless the amount of collapsed baryons, at scales larger than $\sim 10-20\; h^{-1}$Mpc the matter density should closely follow the statistics of a moderately in-homogeneous  Gaussian field.

\subsection{The $vv$ term}

Let us first consider the case where the electron density or the halo number density are {\em constant}. This considerably simplifies Equation (\ref{eq:xips}), since $n_{h,\vk-\vq} = {\bar n}_h\delta^D(\vk-\vq)$ (with $\delta^D$ denoting the Dirac delta):
\begin{equation}
\langle \xi_{\vk_1} \xi_{\vk_2}^* \rangle^{vv} = {\bar n}_h^2 \langle \beta_{\vk_1} \beta_{\vk_2}^*\rangle =
{\bar n}_h^2  \left( 2\pi \right)^3\delta^D (\vk_1-\vk_2)  {\cal D}_{v,1} {\cal D}_{v,2} \frac{P_m(k_1}{k_1^2} 
 (\kh_1\cdot\n_1)(\kh_1\cdot\n_2) \;W_{gas,\vk_1}(r_1) W_{gas,\vk_1}^*(r_2).
\label{eq:xipsvv1}
\end{equation}
The factors ${\cal D}_{v,i}$ denote the peculiar velocity time dependent growth factors along the the $i$-th line of sight. We are using the vorticity free linear theory expression for the Fourier modes of the peculiar velocity, $\vv_{\vk} = i {\cal D}_v \kh \; W_{v,\vk} \delta_{\vk} / k$ with $\kh$ the unit vector along $\vk$. We assign to velocities a Fourier window function $W_{v,\vk}$ that smooths the contribution from scales smaller than a halo if we are referring to a halo's peculiar velocity. In practice we set it practically equal to the gas window function $W_{gas, \vk}$, which in general depends on the halo's mass and age. The particular choice of the velocity window function is not critical, since most of the anisotropy is coming from large scales. The matter power spectrum at present is denoted by $P_m(k)$. After introducing this in Equation (\ref{eq:c121}), one obtains
\[
C(\n_1,\n_2) = \int dr_1 \; a(r_1) \sigma_T n_{e,h}(r_1)  {\bar n}_{h,1} \int dr_2 \; a(r_2) \sigma_T n_{e,h}(r_2)\; {\bar n}_{h,2} \int \dthreek\; {\cal D}_{v,1}{\cal D}_{v,2} \; e^{-i\vk\cdot(\vrv_1-\vrv_2)} \frac{P_m(k)}{k^2} \; \times
\]
\begin{equation}
\phantom{xxxxxxxxxxxxxxxxxxxxxxxxxxxxxxxxxxxxxxx}
 \;W_{gas,\vk}(r_1) W_{gas,\vk}^*(r_2)   \;W_{v,\vk}(r_1) W_{v,\vk}^*(r_2)(\n_1\cdot\kh)(\n_2\cdot\kh),
\label{eq:c121_v1}
\end{equation}
which, after expanding the plane wave on spherical Bessel functions
\begin{equation}
e^{-i\vk\cdot\vrv} = \sum_l (-i)^l (2l+1) P_l(\kh\cdot\n) j_l(kr),
\label{eq:expexp}
\end{equation}
and defining ${\dot \tau}_i \equiv a(r_i) \sigma_T {\bar n}_i$ reads like 
\[
C(\n_1,\n_2) =\sum_{l,l'} \int \dthreek\; \int dr_1 \dot{\tau}_1 {\bar n}_{h,1}{\cal D}_{v,1}W_{gas,\vk}(r_1)W_{v,\vk}(r_1)j_l(kr_1) \;  \int dr_2 \; \dot{\tau}_2{\bar n}_{h,2}{\cal D}_{v,2} W_{gas,\vk}^*(r_2)  W_{v,\vk}^*(r_2)j_{l'}(kr_2) \;   \times
\]
\begin{equation}
\phantom{xxxxxxxxxxxxxxxxxxxxxxxxxxxxxxxxxxxxxxx}
\; \frac{P_m(k)}{k^2} P_l (\n_1\cdot\kh)(\n_1\cdot\kh)\; P_{l'}(\n_2\cdot\kh)(\n_2\cdot\kh) \;(2l+1)(2l'+1)(-i)^{l-l'}.
\label{eq:c121_vv2}
\end{equation}
We next define the line of sight integrals 
\begin{equation}
{\cal I}_{l,i}^{vv} \equiv  \int dr_i \dot{\tau}_i {\bar n}_{h,i}{\cal D}_{v,i}W_{gas,\vk}(r_i)W_{v,\vk}(r_i)j_l(kr_i),
\label{eq:defIvv}
\end{equation}
and apply the recurrence relation
\begin{equation}
\mu P_l(\mu ) = \frac{1}{2l+1}\left( lP_{l-1} (\mu) + (l+1)P_{l+1}(\mu ) \right).
\label{eq:plrel}
\end{equation}
Finally, we perform the integral over the angles of $\vk$. We choose $\n_1$ as the polar axis for $\vk$, so we need to remove the dependence of Legendre polynomials on $\n_2\cdot\kh$. We use the  Legendre function addition theorem, that states that 
\begin{equation}
P_l(\n_2\cdot\kh ) = P_l(\kh\cdot \n_1) P_l(\n_1\cdot \n_2) + 2\sum_{m=1}^l \frac{(l-m)!}{(l+m)!}\; \cos{m\phi_{\vk}}\; P_l^{m} (\kh\cdot \n_1) P_l^m (\n_1\cdot \n_2).
\label{eq:legadth}
\end{equation}
The azimuthal angle of $\vk$ (here denoted as $\phi_{\vk}$) does not appear anywhere else in the integral, and therefore none of the terms in the sum proportional to $\cos m\phi_{\vk}$ contribute to it. These manipulations yield
\[
C(\n_1,\n_2) =\sum_{l} \frac{2l+1}{4\pi}\; P_l(\n_1\cdot \n_2) \; C_l^{vv}\;
=\sum_{l} \frac{2l+1}{4\pi}\; P_l(\n_1\cdot \n_2) \; \left( \frac{2}{\pi} \right) 
\int dk\;k^2 \; \frac{P_m(k)}{k^2} \;\times
\]
\begin{equation}
\phantom{xxxxxx}
\biggl( \frac{(l+1)^2}{(2l+1)^2} {\cal I}^{vv}_{l+1,1}{\cal I}^{vv}_{l+1,2} - \frac{(l+1)l}{(2l+1)^2} {\cal I}^{vv}_{l+1,1}{\cal I}^{vv}_{l-1,2} 
- \frac{(l+1)l}{(2l+1)^2} {\cal I}^{vv}_{l-1,1}{\cal I}^{vv}_{l+1,2}  + \frac{l^2}{(2l+1)^2} {\cal I}^{vv}_{l-1,2}{\cal I}^{vv}_{l-1,2}
\biggr).
\label{eq:c121_vv3}
\end{equation}
Note that in the limit of high $l$ and ${\cal I}^{vv}_{l,1} \equiv {\cal I}^{vv}_{l,2}$, the angular power spectrum multipoles vanish:
\begin{equation}
\label{eq:c121_vvlimit} 
\lim_{l\rightarrow \infty} C_l^{vv} =  \frac{2}{\pi} \int dk\;k^2 \; \frac{P_m(k)}{k^2} \; ({\cal I}_l^{vv})^2 \; \times 
\biggl( \frac{1}{4} - \frac{1}{4} - \frac{1}{4} + \frac{1}{4} \biggr) = 0.
\end{equation}
For this reason it will be required to compute the integrals ${\cal I}_{l,i}^{vv}$ with high accuracy, or the $C_l^{vv}$-s will not drop to zero at high $l$-s as this limit requires.

\subsection{The Poisson term}

This term observes the Poisson statistics of the number of halos along the line of sight. In this case, the power spectrum of $\xi_{\vk}$ reads like
\[
\langle \xi_{\vk_1} \xi_{\vk_2}^* \rangle^P = \int \dthreequno \dthreeqdos \;  \langle n_{\vk_1-\vq_1} n_{\vk_2-\vq_2}^* \rangle^P \langle \beta_{\vq_1} \beta_{\vq_2}^*\rangle =
  \left( 2\pi \right)^3\delta^D (\vk_1-\vq_1-\vk_2+\vq_2) {\bar n}^2 P_{hh}^P(|\vk_1-\vq_1|)  \;\times 
  \]
\begin{equation}
\phantom{x}  
\;W_{\delta,\vk_1-\vq_1} W_{\delta,\vk_2-\vq_2}^*  \;(2\pi)^3 \delta^D(\vq_1-\vq_2)\;{\cal D}_{v,1} {\cal D}_{v,2} \frac{P_m(q_1)}{q_1^2} 
 (\qh_1\cdot\n_1)(\qh_1\cdot\n_2) \;W_{v,\vq_1} W_{v,\vq_2}^*.
\label{eq:xipsvv1}
\end{equation}
The Poisson halo power spectrum is given by $P_{hh}^P(k) = 1/{\bar n}$, i.e., the inverse of the average halo number density. The Fourier density window functions  $W_{\delta,\vk}$ again limit the contribution of the halo power spectrum from scales much smaller than the halo size. They were approximated by Gaussian of characteristic size the halo virial radius. With this, the angular correlation function becomes
\begin{equation}
C(\n_1,\n_2) = \int dr_1 \; {\dot \tau}_1 \int dr_2 \; {\dot \tau}_2\;\;{\bar n}_{h,1} \int \dthreek\; \; e^{-i\vk\cdot(\vrv_1-\vrv_2)} \;|W_{gas,\vk}|^2  \int \dthreequno\; {\cal D}_{v,1}{\cal D}_{v,2}\; \frac{P_m(q_1)}{q_1^2} 
 (\qh_1\cdot\n_1)(\qh_1\cdot\n_2) \;|W_{v,\vq_1}|^2|W_{\delta,\vk-\vq_1}|^2.
\label{eq:c121_p1}
\end{equation}
Let us next focus on the integral over $\vq_1$, which shall be denoted by ${\cal K}$:
\begin{equation}
{\cal K} \equiv \int \dthreequno \; |W_{v,\vq_1}|^2|W_{\delta,\vk-\vq_1}|^2\; \frac{P_m(q_1)}{q_1^2} 
 (\qh_1\cdot\n_1)(\qh_1\cdot\n_2) \;
\label{eq:jpdef}
\end{equation}

 Let us take the polar axis for $\vq_1$ to be along $\kh$. In order to perform the angular integrals of the dot products present in Equation (\ref{eq:jpdef}), we have to switch these products $ (\qh_1\cdot\n_1)$, $(\qh_1\cdot\n_2)$ for others of the type  $ (\qh_1\cdot\kh)$ times other dot products independent of $\qh$. For that, we again make use of the Legendre function addition theorem:
\begin{eqnarray}
\label{eq:dotpp10}
(\qh_1\cdot\n_1) &= & (\qh_1\cdot\kh)(\n_1\cdot\kh) + \cos \phi_1 P_1^1(\qh_1\cdot\kh) P_1^1(\n_1\cdot\kh ), \\
(\qh_1\cdot\n_2) &= & (\qh_1\cdot\kh)(\n_2\cdot\kh) + \cos \phi_2 P_1^1(\qh_1\cdot\kh) P_1^1(\n_2\cdot\kh ). 
\label{eq:dotpp11}
\end{eqnarray}
\vspace{.2cm}
The angles $\phi_{1,2}$ refer to the angles between the planes containing the vector pairs ($\kh$,$\qh_1$) and ($\kh$,$\n_1$), and  ($\kh$,$\qh_1$) and ($\kh$,$\n_2$), respectively. In this reference system where the polar axis is given by $\kh$, the vectors $\qh_1$, $\n_1$ and $\n_2$ have each an azimuthal angle on the plane normal to $\kh$. We shall take the origin for this angle such that the azimuthal angle of $\n_1$ is zero, and write $\phi_2 = \phi_1 + \delta \phi_{21}$.  According to the last two equations, the product $ (\qh_1\cdot\n_1)(\qh_1\cdot\n_2)$
reads
\[
 (\qh_1\cdot\n_1)(\qh_1\cdot\n_2) = (\qh_1\cdot\kh)^2(\n_1\cdot\kh)(\n_2 \cdot\kh)+ 
 \biggl( P_1^1(\qh_1\cdot\kh) \biggr)^2 \biggl[  \cos \phi_1 P_1^1(\n_1\cdot\kh ) \cos \phi_2 P_1^1(\n_2\cdot\kh ) \biggr] +
 \]
\begin{equation} 
\phantom{xxxxxx}
  \cos \phi_1 P_1^1(\qh_1\cdot\kh) P_1^1(\n_1\cdot\kh ) (\qh_1\cdot\kh)(\n_2\cdot\kh) +  \cos \phi_2 P_1^1(\qh_1\cdot\kh) P_1^1(\n_2\cdot\kh )(\qh_1\cdot\kh)(\n_1\cdot\kh)
\label{eq:dotpp}
\end{equation}
The last two terms proportional to $\cos \phi_{1,2}$ will give no contribution once the integration on the azimuthal angle of $\vq$ ($\phi_1$) is done, and therefore can be neglected. Since $\phi_2 = \phi_1 + \delta \phi_{21}$, then $\cosÊ\phi_2 = \cos \phi_1 \cos \delta\phi_{21} - \sin \phi_1 \sin \delta \phi_{21}$. The product of $\cos \phi_1\cos\phi_2$ will therefore give rise to a term proportional to $\cos \phi_1  \sin \phi_1$ that again gives zero contribution to the integral over the azimuthal angle of $\vq_1$ ($\phi_1$).  Thus, effectively, we are left with the term proportional to $\cos^2 \phi_1$:
\begin{equation} 
 (\qh_1\cdot\n_1)(\qh_1\cdot\n_2) = (\qh_1\cdot\kh)^2(\n_1\cdot\kh)(\n_2 \cdot\kh)+ 
 \biggl( P_1^1(\qh_1\cdot\kh) \cos\phi_1\biggr)^2 \biggl[   \cos \delta \phi_{21} P_1^1(\n_1\cdot\kh )  P_1^1(\n_2\cdot\kh ) \biggr] .
\label{eq:dotpp2}
\end{equation}
We again invoke the Legendre function addition theorem  on the triad formed by the vectors $\kh$, $\n_1$ and $\n_2$ in order to manipulate the expression within the square brackets of the last term on the rhs of the last equation. It now becomes
\begin{equation} 
 (\qh_1\cdot\n_1)(\qh_1\cdot\n_2) = (\qh_1\cdot\kh)^2(\n_1\cdot\kh)(\n_2 \cdot\kh)+ 
 \biggl( P_1^1(\qh_1\cdot\kh) \cos\phi_1\biggr)^2 \biggl[   (\n_1\cdot \n_2) - (\n_1\cdot\kh)(\n_2\cdot\kh) \biggr] .
\label{eq:dotpp3}
\end{equation}
This expression can be plugged into Equation (\ref{eq:jpdef}):
\[
{\cal K} = \int \frac{q_1^2dq_1}{(2\pi)^3} \int_0^{2\pi} d\phi_1 \int_{-1}^1 d\mu_{q_1}  |W_{v,\vq_1}|^2|W_{\delta,\vk-\vq_1}|^2\; \frac{P_m(q_1)}{q_1^2}  \biggl[  \cos^2 \phi_1 |P_1^1(\mu_{q_1})|^2 (\n_1\cdot \n_2)\; +   
\]
\begin{equation}
\phantom{xxxxxxxxxxxx}
\left( \mu_{q_1}^2 -\cos^2 \phi_1 |P_1^1(\mu_{q_1})|^2\right)(\n_1\cdot\kh)(\n_2\cdot\kh) \biggr],
\label{eq:jpdef2}
\end{equation}
where $\mu_{q_1} \equiv \qh_1\cdot\kh$. We next assume that, on average, halos are spherically symmetric, so that $W_{\delta, \vk} = W_{\delta, k}$ and $W_{v,\vk} = W_{v,k}$. This simplifies the last equation,
\begin{equation}
{\cal K} =  \biggl[ {\cal J}_1(k)  (\n_1\cdot \n_2) - {\cal J}_2(k) (\n_1\cdot\kh)(\n_2\cdot\kh) \biggr],
\label{eq:jpdef3}
\end{equation}
with ${\cal J}_1(k)$ defined as
\begin{equation}
{\cal J}_1(k) \equiv \int \frac{q_1^2dq_1}{(2\pi)^3} \; \pi \int_{-1}^1 d\mu_{q_1}\;  |W_{v,q_1}|^2|W_{\delta,|\vk-\vq_1|}|^2\; \frac{P_m(q_1)}{q_1^2},
\label{eq:j1pdef1}
\end{equation}
and ${\cal J}_2(k)$ as
\begin{equation}
{\cal J}_2(k) \equiv \int \frac{q_1^2dq_1}{(2\pi)^3} \; \int_0^{2\pi} d\phi_1 \int_{-1}^1 d\mu_{q_1}\; |W_{v,q_1}|^2|W_{\delta,|\vk-\vq_1|}|^2\; \frac{P_m(q_1)}{q_1^2} \biggl[\mu^2_{q_1} - |P_1^1(\mu_{q_1})|^2 \cos ^2 \phi_1\biggr].
\label{eq:j2pdef1}
\end{equation}

This also simplifies the expression for the angular correlation function (Equation \ref{eq:c121_p1}):
\begin{equation}
C(\n_1,\n_2) = \int dr_1 \; {\dot \tau}_1 \int dr_2 \; {\dot \tau}_2\;\;{\bar n}_{h,1}  \;{\cal D}_{v,1}{\cal D}_{v,2}\; \int \dthreek\; \; e^{-i\vk\cdot(\vrv_1-\vrv_2)} \;|W_{gas,\vk}|^2  \biggl[{\cal J}_1(k)   (\n_1\cdot \n_2) + (\n_1\cdot\kh)(\n_2\cdot\kh) {\cal J}_2(k)\biggr] .
\label{eq:c121_p2}
\end{equation}
Let us split it into two contributions, according to their angular dependence. The first contribution $C^P_A(\n_1,\n_2)$ contains the term $  (\n_1\cdot \n_2)$, the second one ($C^P_B(\n_1,\n_2)$) contains $(\n_1\cdot\kh)(\n_2\cdot\kh) $. The procedure now is very similar to that used for the $vv$ term. We first expand the plane wave as in Equation (\ref{eq:expexp}), and define the line of sight integrals ${\cal I}^P_{l,i}$ as
\begin{equation}
{\cal I}^P_{l,i} \equiv \int dr_i \; {\dot \tau}_i\;\sqrt{{\bar n}_{h,i}}\; W_{gas,\vk} {\cal D}_{v,i} j_l(kr_i).
\label{eq:defIlp}
\end{equation}

\subsubsection{The $C^P_A(\n_1,\n_2)$ part}
The contribution from $C^P_A(\n_1,\n_2)$ then reads
\begin{equation}
C^P_A(\n_1,\n_2) = \sum_{l,l'} (2l+1)(2l'+1)(-i)^{l-l'} \int \dthreek\;{\cal J}_1(k) \;{\cal I}^P_{l,1} {\cal I}^P_{l,2}\; P_l(\n_1\cdot\kh) P_{l'}(\n_2\cdot\kh) (\n_1\cdot \n_2).
\label{eq:c121_pA1}
\end{equation}
The addition theorem applied on $ P_{l'}(\n_2\cdot\kh) $ and the integral on the azimuthal angle (that removes terms of the type $\cos m\phi_{\vk}$) leave the last expression as 
\begin{equation}
C^P_A(\n_1,\n_2) = \sum_{l,l'} (2l+1)(2l'+1)(-i)^{l-l'} \int \frac{k^2dkd\mu_{\vk}}{(2\pi)^2}\;{\cal J}_1(k) \;{\cal I}^P_{l,1} {\cal I}^P_{l,2}\; P_l(\n_1\cdot\kh) P_{l'}(\n_1\cdot\kh) P_{l'}(\n_1\cdot \n_2) (\n_1\cdot \n_2).
\label{eq:c121_pA2}
\end{equation}
Finally, the recurrence relation (\ref{eq:plrel}) yields the final expression for $C^P_A(\n_1,\n_2)$, after integrating in $\mu_{\vk}$:
\begin{equation}
C^P_A(\n_1,\n_2) = \sum_{l} \frac{2l+1}{4\pi} P_l(\n_1\cdot \n_2) C_{l,A}^{P} =  \sum_{l} \frac{2l+1}{4\pi} P_l(\n_1\cdot \n_2) \; \left( \frac{2}{\pi}\right) \int k^2dk\;{\cal J}_1(k) \biggl( \frac{l+1}{2l+1}{\cal I}^P_{l+1,1} {\cal I}^P_{l+1,2}  +  \frac{l}{2l+1}{\cal I}^P_{l-1,1} {\cal I}^P_{l-1,2}\biggr).
\label{eq:c121_pA3}
\end{equation}
In this case, this contribution does not vanish at high $l$:
\begin{equation}
\lim_{l\rightarrow \infty} C_{l,A}^{P} =  \frac{2}{\pi} \int k^2dk\;{\cal J}_1(k) |{\cal I}^P_{l}|^2.
\label{eq:limic121_pA}
\end{equation}

\subsubsection{The $C^P_B(\n_1,\n_2)$ part}

This contribution is written like
\begin{equation}
C^P_B(\n_1,\n_2) = -\sum_{l,l'} (2l+1)(2l'+1)(-i)^{l-l'} \int \dthreek\;{\cal J}_2(k) \;{\cal I}^P_{l,1} {\cal I}^P_{l,2}\; P_l(\n_1\cdot\kh) P_{l'}(\n_2\cdot\kh) (\n_1\cdot\kh)(\n_2\cdot\kh), 
\label{eq:c121_pB1}
\end{equation}
which, after using the Legendre polynomial recurrence relation (Equation \ref{eq:plrel}) and integrating on the angles, yields
\[
C^P_B(\n_1,\n_2) = -\sum_{l} \frac{2l+1}{4\pi} P_l(\n_1\cdot \n_2) C_{l,B}^{P} =  -\sum_{l} \frac{2l+1}{4\pi} P_l(\n_1\cdot \n_2) \; \left( \frac{2}{\pi}\right) \int k^2dk\;{\cal J}_2(k)\times
\]
\begin{equation}
 \biggl(    \frac{(l+1)^2}{(2l+1)^2} {\cal I}^{P}_{l+1,1}{\cal I}^{P}_{l+1,2} - \frac{(l+1)l}{(2l+1)^2} {\cal I}^{P}_{l+1,1}{\cal I}^{P}_{l-1,2} 
- \frac{(l+1)l}{(2l+1)^2} {\cal I}^{P}_{l-1,1}{\cal I}^{P}_{l+1,2}  + \frac{l^2}{(2l+1)^2} {\cal I}^{P}_{l-1,2}{\cal I}^{P}_{l-1,2}  \biggr).
\label{eq:c121_pB2}
\end{equation}
As it was the case for the $vv$ term (which showed an identical dependence on $\n_1$, $\n_2$), this contribution will vanish at high $l$ if ${\cal I}_{l,1}^{P,B} \equiv {\cal I}_{l,2}^{P,B}$:
\begin{equation}
\label{eq:c121_Plimit} 
\lim_{l\rightarrow \infty} C_{l,B}^{P} =  \frac{2}{\pi} \int dk\;k^2 \; {\cal J}_2(k)\; ({\cal I}_l^{P})^2 \; \times 
\biggl( \frac{1}{4} - \frac{1}{4} - \frac{1}{4} + \frac{1}{4} \biggr) = 0.
\end{equation}
Therefore, we find that $\lim_{l\rightarrow \infty} C_l^{P} = C_{l,A}^{P}$.
\vspace{1.cm}

We have concluded the computation of the linear terms, and we next address the computation of the second-order terms. In this Section, the electron density will be assumed to trace the dark matter density field. This means that, when looking at the kSZ generated in halos, the number density of halos will trace the local, large scale, dark matter density field, in such a way that fluctuations in the halo number density will be a {\em biased} mapping  of the fluctuation of dark matter density (and this bias factor was approximated by that of  \citet{mowhite96}). I.e., $n_{h,\vk} = b{\cal D}_{\delta}\delta_{\vk} W_{\delta,\vk}$, with $b$ the bias factor, $n_{h,\vk}$ the $\vk$ Fourier mode for the halo number density distribution, $\delta_{\vk}$ the Fourier mode of the dark matter density contrast {\em at present}, and ${\cal D}_{\delta}$ its linear growth factor. When considering the OV generated by the IGM on the large scales, this assumption is expressed as  $n_{e,\vk} = {\bar n}_e(r){\cal D}_{\delta}\; \delta_{\vk}$ with ${\bar n}_e(r)$ the background electron number density.

\subsection{The $v_1d_1-v_2d_2$ term}

This term observes the contribution of the configuration
\begin{equation}
\langle \xi_{\vk_1}\xi_{\vk_2}^{*}\rangle^{v_1d_1-v_2d_2} \equiv \int \dthreequno \dthreeqdos \; \langle \beta_{\vq_1} n_{h,\vk_1-\vq_1}\rangle \langle \beta^*_{\vq_2} n^*_{h,\vk_2-\vq_2} \rangle.
\label{eq:xiAdef}
\end{equation}
As we show next, this contribution vanishes. For a real field $A(\vx )$, we have that $A_{\vk} = A_{-\vk}^*$, so the last equation can be rewritten as
\begin{equation}
\langle \xi_{\vk_1}\xi_{\vk_2}^{*}\rangle^{v_1d_1-v_2d_2}  \equiv \int \dthreequno \dthreeqdos \; (2\pi)^3 \delta^D(\vk_1) \frac{{\cal D}_{v,1}}{q_1} P_m(q_1) (\qh_1\cdot \n_1)\; 2\pi)^3 \delta^D(\vk_2) \frac{{\cal D}_{v,2}}{q_2} P_m(q_2) (\qh_2\cdot \n_2).
\label{eq:xiAdef1}
\end{equation}
If we choose $\n_i$ as the polar angle for $\vq_i$ ($i=1,2$), then $\mu_i \equiv (\qh_i\cdot \n_i)$ and we end up with an integral of the type
\begin{equation}
\int \frac{q_i^2dq_i}{(2\pi)^3} d\phi_{i} \frac{P_m(q_i)}{q_i}\int_{-1}^1d\mu_{i} \; \mu_{i} = 0.
\label{eq:intvanish}
\end{equation}
Therefore, this term provides no contribution and will be ignored hereafter.

\subsection{The $vv-dd$ term}

We next compute the contribution of the configuration
\begin{equation}
\langle \xi_{\vk_1}\xi_{\vk_2}^{*}\rangle^{vv-dd} \equiv \int \dthreequno \dthreeqdos \; \langle \beta_{\vq_1} \beta_{\vq_2}^*\rangle \langle n_{h,\vk_1-\vq_1} n^*_{h,\vk_2-\vq_2} \rangle
\label{eq:xiBdef}
\end{equation}
within the two-point angular correlation function
\begin{equation}
C^{vv-dd}(\n_1,\n_2) = \int dr_1 \; a(r_1) \sigma_T n_{e,h}(r_1) \int dr_2 \; a(r_2) \sigma_T n_{e,h}(r_2)\; \dthreekuno \dthreekdos\; e^{-i\vk_1\cdot\vrv_1 + i\vk_2\cdot\vrv_2}W_{gas,\vk_1} W_{gas,\vk_2}^* \langle \xi_{\vk_1} \xi_{\vk_2}^*\rangle^{vv-dd}.
\label{eq:c12Bdef}
\end{equation}
The ensemble averages of Equation (\ref{eq:xiBdef}) yield
\[
\langle \xi_{\vk_1}\xi_{\vk_2}^{*}\rangle^{vv-dd} = \int \dthreequno \dthreeqdos \;  (2\pi)^3\delta^D(\vq_1-\vq_2){\cal D}_{v,1}{\cal D}_{v,2} \frac{P_m(q_1)}{q_1^2} W_{v,\vq_1} {\bar n}_{h,1}W_{v,\vq_2}^*  {\bar n}_{h,2}\;  (\qh_1 \cdot \n_1) (\qh_2 \cdot \n_2) \;\times 
\]
\begin{equation}
\phantom{xxxxxxxxxxxxxxxxxxxxxxxxxxxx}
(2\pi)^3\delta^D(\vk_1-\vq_1 - \vk_2 +\vq_2) {\cal D}_{\delta,1}b_1 {\cal D}_{\delta,2} b_2 P_m(\vk_1-\vq_1) W_{\delta,\vk_1-\vq_1 }W_{\delta,\vk_2-\vq_2 }^* .
\label{eq:xiB2}
\end{equation}
The presence of these Dirac deltas significantly simplifies Equation (\ref{eq:c12Bdef}):
\[
C^{vv-dd}(\n_1,\n_2) = \int \dthreek e^{-i\vk(\vrv_1-\vrv_2)} \;\int dr_1 \; {\dot \tau}_1W_{gas,\vk}  {\bar n}_{h,1}b_1{\cal D}_{v,1}{\cal D}_{\delta,1}b_1  \int dr_2 \;  {\dot \tau}_2 W_{gas,\vk}^*  {\bar n}_{h,2}{\cal D}_{v,2}{\cal D}_{\delta,2}b_2 \; \times 
\]
\begin{equation}
\phantom{xxxxxxxxxxxxxxxxxxxxxxxxxxxx}
\int \dthreequno\; \frac{P_m(q_1)}{q_1^2} |W_{v,\vq_1}|^2  (\qh_1 \cdot \n_1) (\qh_1 \cdot \n_2) \; P_m(\vk-\vq_1) |W_{\delta,\vk-\vq_1 }|^2.
\label{eq:c12B1}
\end{equation}
There  are functions inside the integral over $\vq_1$ which depend upon $\vk-\vq_1$. Just as for the Poisson term, this suggests using $\kh$ as the polar axis for the angular integration of $\vq_1$, $\mu_{q_1} \equiv \kh\cdot \qh_1$. Thus we have to apply the Legendre function addition theorem and follow the same procedure as in Equations (\ref{eq:dotpp10} -- \ref{eq:dotpp3}). After introducing the usual expansion of the plane waves, this gives
\[
C^{vv-dd}(\n_1,\n_2) = \sum_{l,l'}(-i)^{l-l'}\int \dthreek  {\cal I}_{l,1}^{vv-dd} {\cal I}_{l,2}^{vv-dd}  P_l(\n_1\cdot \kh) P_{l'}(\n_2\cdot \kh) \;  
\int \dthreequno\; \frac{P_m(q_1)}{q_1^2} |W_{v,\vq_1}|^2   P_m(\vk-\vq_1) |W_{\delta,\vk-\vq_1 }|^2 \;\times
\]
\begin{equation}
\phantom{xxxxxxxxxx}
\biggl[ 
(\qh_1\cdot \kh_1)^2 (\n_1\cdot \kh) (\n_2\cdot \kh) + \biggl( \cos \phi_1 P_1^1(\kh\cdot \qh_1)\biggr) ^2 \biggl( \n_1\cdot \n_2 - (\n_1\cdot \kh)(\n_2\cdot \kh)\biggr)
\biggr] (2l+1)(2l'+1),
\label{eq:c12B2}
\end{equation}
where the ${\cal I}_{l,i}^{vv-dd} \equiv \int dr_i \; j_l(kr_i) {\dot \tau}_i  {\bar n}_{h,i}W_{gas,\vk} b_i{\cal D}_{v,i}{\cal D}_{\delta,i} $ are the usual line of sight integrals. The integration on the azimuthal angle of $\vq_1$ can be done trivially, but the integration on its polar angle must be done numerically. These angular integrations introduce two $k$-dependent functions, ${\cal A}(k)$ and ${\cal B}(k)$:
\[
C^{vv-dd}(\n_1,\n_2) = \sum_{l,l'}(-i)^{l-l'}\int \dthreek  {\cal I}_{l,1}^{vv-dd} {\cal I}_{l',2}^{vv-dd}  P_l(\n_1\cdot \kh) P_{l'}(\n_2\cdot \kh) \;  
\int  \frac{q_1^2 dq_1 d\mu_{q_1}}{(2\pi)^3}\; \frac{P_m(q_1)}{q_1^2} |W_{v,\vq_1}|^2   P_m(\vk-\vq_1) |W_{\delta,\vk-\vq_1 }|^2 \;\times
\]
\[
\phantom{xxxxxxxxxx}
(2\pi)\;\biggl[  (\n_1\cdot \kh) (\n_2\cdot \kh) \biggl( (\kh\cdot\qh_1)^2 - \frac{1}{2} \left( P_1^1(\kh\cdot\qh_1)\right)^2 \biggr)+ \frac{1}{2}\left( P_1^1(\kh\cdot\qh_1)\right)^2 (\n_1\cdot \n_2) \biggr] (2l+1)(2l'+1)= 
\]
\[
 \sum_{l,l'}(-i)^{l-l'}\int \dthreek  {\cal I}_{l,1}^{vv-dd} {\cal I}_{l',2}^{vv-dd}  P_l(\n_1\cdot \kh) P_{l'}(\n_2\cdot \kh) \;  
\int \frac{q_1^2 dq_1}{(2\pi)^3}\; \frac{P_m(q_1)}{q_1^2} |W_{v,\vq_1}|^2   P_m(\vk-\vq_1) |W_{\delta,\vk-\vq_1 }|^2 \;\times
\]
\begin{equation}
\biggl[ {\cal A}(k) (\n_1\cdot \kh) (\n_2\cdot \kh) + {\cal B}(k)  (\n_1\cdot \n_2)   \biggr] (2l+1)(2l'+1).
\label{eq:c12B3}
\end{equation}
These two functions will give rise to two different contributions for the $vv-dd$ term, $C^{vv-dd}_{A} (\n_1,\n_2)$ and $C^{vv-dd}_{ B} (\n_1,\n_2)$.

\subsubsection{The $C^{vv-dd}_{ A} (\n_1,\n_2)$ contribution}

As already done for the $vv$ term and one of the two contributions for the Poisson term, we first apply the Legendre polynomial recurrence relation (Equation \ref{eq:plrel}) in order to express all the dependence on $\kh\cdot \n_1$ and $\kh\cdot \n_2$ as arguments of Legendre polynomials:
\begin{equation}
C_A^{vv-dd}(\n_1,\n_2) = \sum_{l,l'}(-i)^{l-l'}\int \dthreek  {\cal I}_{l,1}^{vv-dd} {\cal I}_{l',2}^{vv-dd} \;
\biggl[l P_{l-1}(\n_1\cdot \kh) + (l+1)P_{l+1}(\n_1\cdot \kh) \biggr] \biggl[ l' P_{l'-1}(\n_2\cdot \kh) + (l'+1)P_{l'+1}(\n_2\cdot \kh) \biggr] {\cal A}(k).
\label{eq:c12Ba1}
\end{equation}
The next step uses the Legendre function addition theorem (Equation \ref{eq:legadth}) to rewrite Legendre polynomials whose argument is $\kh\cdot \n_2$ in terms of Legendre functions of arguments $\kh\cdot \n_1$ and $\n_1 \cdot \n_2$. The integral on the azimuthal angle of $\kh$ removes all terms proportional to $\cos m\phi_{\vk}$, so in practice we simply substitute the Legendre polynomials $P_l(\kh\cdot \n_2)$ by the product $P_l(\kh\cdot \n_2) P_l(\n_1\cdot\n_2)$. The polar angular integral of $\kh$ does the rest:
\[
C^{vv-dd}_A(\n_1,\n_2) = \sum_l \frac{2l+1}{4\pi} P_l(\n_1\cdot\n_2) C_{l,A}^{vv-dd} = \sum_l \frac{2l+1}{4\pi} P_l(\n_1\cdot\n_2) \; \times
\]
\begin{equation}
\phantom{xxxxx}
\frac{2}{\pi}\int k^2dk\; {\cal A}(k) \biggl[ {\cal I}_{l+1,1}^{vv-dd} \biggl( \frac{(l+1)^2}{(2l+1)^2}{\cal I}_{l+1,2}^{vv-dd} - \frac{l(l+1)}{(2l+1)^2}{\cal I}_{l-1,2}^{vv-dd}\biggr) + {\cal I}_{l-1,1}^{vv-dd} \biggl( - \frac{l(l+1)}{(2l+1)^2} {\cal I}_{l-1,2}^{vv.dd} +  \frac{l^2}{(2l+1)^2} {\cal I}_{l-1,2}^{vv-dd}\biggr)
\biggr].
\label{eq:c12Ba2}
\end{equation}
As for every term with the angular dependence of the type $ (\n_1\cdot \kh) (\n_2\cdot \kh)$, its contribution vanishes in the limit of $l\rightarrow \infty$ and ${\cal I}_{l,1}^{vv-dd} \equiv {\cal I}_{l,2}^{vv-dd}$:
\begin{equation}
\lim_{l\rightarrow \infty} C_{l,A}^{vv-dd} = \frac{2}{\pi} \int dk\;k^2 \; {\cal A}(k) \; ({\cal I}_l^{vv-dd})^2 \; \times 
\biggl( \frac{1}{4} - \frac{1}{4} - \frac{1}{4} + \frac{1}{4} \biggr) = 0.
\label{eq:limCBa}
\end{equation}

\subsubsection{The $C^{vv-dd}_{ B} (\n_1,\n_2)$ contribution}

In this case we again have to use the addition theorem of Equation (\ref{eq:legadth}) to obtain
\begin{equation}
C^{vv-dd}_B(\n_1,\n_2) = \sum_{l,l'}(-i)^{l-l'}(2l+1)(2l'+1)\int \dthreek  {\cal I}_{l,1}^{vv-dd} {\cal I}_{l',2}^{vv-dd} \;
{\cal B}(k) \frac{2\delta^K_{l,l'}}{2l+1} P_{l'}(\n_1\cdot\n_2) (\n_1\cdot\n_2),
\label{eq:c12Bb1}
\end{equation}
where $\delta^K_{i,j}$ represents the Kronecker delta ($\delta^K_{i,j} = 1$ if $i=j$ and $\delta^K_{i,j} = 0$ otherwise). Finally, an application of the relation (\ref{eq:plrel}) gives
\begin{equation}
C^{vv-dd}_A(\n_1,\n_2) = \sum_{l} \frac{2l+1}{4\pi} P_l(\n_1\cdot\n_2) C_{l,B}^{vv-dd} = 
\sum_{l} \frac{2l+1}{4\pi} P_l(\n_1\cdot\n_2)\; \frac{2}{\pi}\int k^2dk\; {\cal B}(k) \biggl[ {\cal I}_{l+1,1}^{vv-dd} 
{\cal I}_{l+1,2}^{vv-dd} \frac{l+1}{2l+1} +  {\cal I}_{l-1,1}^{vv-dd} 
{\cal I}_{l-1,2}^{vv-dd} \frac{l}{2l+1} 
\biggr] .
\label{eq:c12Bb2}
\end{equation}
Hence, this term gives most of the contribution of $C_l^{vv-dd}$ at high $l$-s:
\begin{equation}
\lim_{l\rightarrow \infty} C_{l,B}^{vv-dd} = \frac{2}{\pi} \int dk\;k^2 \; {\cal B}(k) \; ({\cal I}_l^{vv-dd})^2 = \lim_{l\rightarrow \infty} C_{l}^{vv-dd}.
\label{eq:limCBb}
\end{equation}

\subsection{The $vd-vd$ term}

In this final section, we compute the contribution coming from the configuration
\begin{equation}
\langle \xi_{\vk_1}\xi_{\vk_2}^{*}\rangle^{vd-vd} \equiv \int \dthreequno \dthreeqdos \; \langle \beta_{\vq_1} n_{h,\vk_2-\vq_2}^*\rangle \langle n_{h,\vk_1-\vq_1} \beta^*_{\vq_2} \rangle
\label{eq:xiCdef}
\end{equation}
within the two-point angular correlation function
\begin{equation}
C^{vd-vd}(\n_1,\n_2) = \int dr_1 \; a(r_1) \sigma_T n_{e,h}(r_1) \int dr_2 \; a(r_2) \sigma_T n_{e,h}(r_2)\; \dthreekuno \dthreekdos\; e^{-i\vk_1\cdot\vrv_1 + i\vk_2\cdot\vrv_2}W_{gas,\vk_1} W_{gas,\vk_2}^* \langle \xi_{\vk_1} \xi_{\vk_2}^*\rangle^{vd-vd}.
\label{eq:c12Cdef}
\end{equation}
This time, the computation of the ensemble averages of Equation (\ref{eq:xiCdef}) is slightly more complex. We first obtain, after canceling the integral on $\vk_2$ by using one of the Dirac deltas,
\[
\langle \xi_{\vk_1}\xi_{\vk_2}^{*}\rangle^{vd-vd} = \int \dthreequno (2\pi)^3\delta^D(\vq_1+\vk_1 -\vq_1 - \vk_2)
 {\bar n}_{h,2}b_2 {\cal D}_{\delta,2} {\cal D}_{v,1} W_{v,\vq_1} W^*_{\delta,\vk_2-\vk_1+\vq_1}(\qh_1\cdot \n_1) \frac{P_m(q_1)}{q_1}  \; \times
\]
\begin{equation}
\phantom{xxxxx}
 {\bar n}_{h,1}b_1 {\cal D}_{\delta,1} {\cal D}_{v,2}\left(\frac{\vk_1-\vq_1}{|\vk_1 - \vq_1|}\cdot \n_2 \right)  W_{\delta,\vk_1-\vq_1} W_{v,\vk_1-\vq_1}^* \;   \frac{P_m(\vk_1-\vq_1)}{|\vk_1 - \vq_1|}.
\label{eq:xiC2}
\end{equation}
Again we have an integral on $\vq_1$ of functions whose argument is $\vk_1-\vq_1$. This means that we have to translate all arguments of the type $(\qh_1\cdot \n_1)$ and $(\qh_1\cdot \n_2)$ onto the quantities $(\qh_1\cdot \kh_1)$ and $(\n_1\cdot\n_2)$ and choose $\kh_1$ as the polar axis for the $\vq_1$ integration. By using Equations (\ref{eq:dotpp10}, \ref{eq:dotpp11}) and neglecting the terms proportional to $\cos \phi_1$, $\cos \phi_2$ (which give no contribution after the integration on the azimuthal angle of $\vq_1$), one readily finds
\[
\langle \xi_{\vk_1}\xi_{\vk_2}^{*}\rangle^{vd-vd} = (2\pi)^3\delta^D(\vk_1 - \vk_2) \int \dthreequno 
 {\bar n}_{h,1} {\bar n}_{h,2}b_1b_2 {\cal D}_{\delta,2} {\cal D}_{v,1} {\cal D}_{\delta,1} {\cal D}_{v,2} W_{v,\vq_1} W^*_{\delta,\vq_1}W_{\delta,\vk_1-\vq_1} W_{v,\vk_1-\vq_1}^* \frac{P_m(q_1)}{q_1}  \frac{P_m(\vk_1-\vq_1)}{|\vk_1 - \vq_1|} \; \times
\]
\begin{equation}
\phantom{xxxxx}
\biggl[ k_1(\kh_1\cdot \n_1)(\kh_1\cdot \n_2) (\qh_1\cdot \kh_1) - q_1\left( (\qh_1\cdot\kh_1)^2(\kh_1\cdot \n_1)(\kh_1\cdot \n_2) + \cos\phi_1 \cos\phi_2|P_1^1(\qh_1\cdot \kh_1|^2 P_1^1(\kh_1\cdot \n_1)P_1^1(\kh_1\cdot \n_1)\right)
\biggr].
\label{eq:xiC3}
\end{equation}
As we did for the Poisson term, we now express $\cos \phi_2$ in terms of $\phi_1$ and $\delta \phi_{21} \equiv \phi_2 - \phi_1$ and make use of the relation $\cos \delta \phi_{21} P_1^1(\kh_1\cdot \n_1) P_1^1(\kh_1\cdot \n_2)  = (\n_1\cdot \n_2) - (\kh_1\cdot \n_1)(\kh_2 \cdot \n_2)$. After integrating in the azimuthal angle of $\vq_1$, this leads to
\[
\langle \xi_{\vk_1}\xi_{\vk_2}^{*}\rangle^{vd-vd} = (2\pi)^3\delta^D(\vk_1 - \vk_2) \int \frac{q_1^2dq_1 d\mu_{q_1}}
{(2\pi)^3} {\bar n}_{h,1} {\bar n}_{h,2}b_1b_2 {\cal D}_{\delta,2} {\cal D}_{v,1} {\cal D}_{\delta,1} {\cal D}_{v,2} W_{v,\vq_1} W^*_{\delta,\vq_1}W_{\delta,\vk_1-\vq_1} W_{v,\vk_1-\vq_1}^* \frac{P_m(q_1)}{q_1}  \frac{P_m(\vk_1-\vq_1)}{|\vk_1 - \vq_1|} \; \times
\]
\begin{equation}
\phantom{xxxxx}
(2\pi)\biggl[ (\kh_1\cdot \n_1)(\kh_1\cdot \n_2) \left( k_1\mu_{q_1} - q_1\mu_{q_1}^2 + \frac{q_1}{2} |P_1^1 (\mu_{q_1})|^2\right) - \frac{q_1}{2} |P_1^1(\mu_{q_1})|^2 (\n_1\cdot \n_2) \biggr].
\label{eq:xiC4}
\end{equation}
We plug this expression into Equation (\ref{eq:c12Cdef}) after expanding the plane wave onto spherical Bessel functions and defining ${\dot \tau}_i \equiv  a(r_i) \sigma_T n_{e,h}(r_i)$ and the line of sight integral
${\cal I}_{l,i}^{vd-vd} \equiv {\cal I}_{l,i}^{vv-dd} = \int dr_i   \; j_l(kr_i) {\dot \tau}_i W_{gas,\vk}  {\bar n}_{h,i}b_i{\cal D}_{v,i}{\cal D}_{\delta,i}$. We obtain
\[
C^{vd-vd}(\n_1,\n_2) = \sum_{l,l'}(-i)^{l-l'}\int \dthreek  {\cal I}_{l,1}^{vd-vd} {\cal I}_{l',2}^{vd-vd}  P_l(\n_1\cdot \kh) P_{l'}(\n_2\cdot \kh) \;  
\int  \frac{q_1^2 dq_1 d\mu_{q_1}}{(2\pi)^3}\; \frac{P_m(q_1)}{q_1}\frac{P_m(\vk-\vq_1)}{|\vk-\vq_1|^2}\; \
(2\pi)\times
\]
\[
\phantom{xxxxx}
\biggl[ (\kh\cdot \n_1)(\kh\cdot \n_2) \left( k_1\mu_{q_1} - q_1\mu_{q_1}^2 + \frac{q_1}{2} |P_1^1 (\mu_{q_1})|^2\right) - \frac{q_1}{2} |P_1^1(\mu_{q_1})|^2 (\n_1\cdot \n_2)
\biggr] (2l+1) (2l'+1) = 
\]
\begin{equation}
\phantom{xx} \sum_{l,l'}(-i)^{l-l'}\int \dthreek  {\cal I}_{l,1}^{vd-vd} {\cal I}_{l',2}^{vd-vd}  P_l(\n_1\cdot \kh) P_{l'}(\n_2\cdot \kh) \;  \; 
(2\pi)\biggl[ (\kh\cdot \n_1)(\kh\cdot \n_2) {\cal C}(k) - {\cal D}(k) (\n_1\cdot \n_2) \biggr] (2l+1)(2l'+1).
\label{eq:c12C2}
\end{equation}
The two functions ${\cal C}(k)$ and ${\cal D}(k)$ introduce two different angular dependences of the correlation function ($(\kh\cdot \n_1)(\kh\cdot \n_2) $ and $(\n_1\cdot \n_2) $, respectively), and give rise to two different contributions to the total $vd-vd$ correlation function, that will be denoted by $C^{vd-vd}_A (\n_1,\n_2)$ and $C^{vd-vd}_B (\n_1,\n_2)$, respectively.

\subsubsection{The $C^{vd-vd}_A (\n_1,\n_2)$ contribution}

As usual, we first use Equation (\ref{eq:plrel}) to remove the factors of the type $xP_l(x)$ and then apply the Legendre function addition theorem to switch arguments of $(\kh\cdot \n_2)$ into arguments of $(\kh\cdot \n_1)$ and  $(\n_1\cdot \n_2)$. All terms proportional to $\cos m\phi_{\vk}$ give no contribution, so under the integral sign we are allowed to replace $P_l(\kh\cdot \n_2)$ by $P_l(\kh\cdot \n_1)P_l(\n_1\cdot \n_2)$. This results in
\[
C^{vd-vd}_A(\n_1,\n_2) = \sum_l \frac{2l+1}{4\pi} P_l(\n_1\cdot\n_2) C_{l,A}^{vd-vd} = \sum_l \frac{2l+1}{4\pi} P_l(\n_1\cdot\n_2) \; \times
\]
\begin{equation}
\phantom{xxxxx}
\frac{2}{\pi}\int k^2dk\; {\cal C}(k) \biggl[ {\cal I}_{l+1,1}^{vd-vd} \biggl( \frac{(l+1)^2}{(2l+1)^2}{\cal I}_{l+1,2}^{vd-vd} - \frac{l(l+1)}{(2l+1)^2}{\cal I}_{l-1,2}^{vd-vd}\biggr) + {\cal I}_{l-1,1}^{vd-vd} \biggl( - \frac{l(l+1)}{(2l+1)^2} {\cal I}_{l-1,2}^{vd-vd} +  \frac{l^2}{(2l+1)^2} {\cal I}_{l-1,2}^{vd-vd}\biggr)
\biggr].
\label{eq:c12Ca1}
\end{equation}
And again, in the limit of high $l$ and ${\cal I}_{l,1}^{vd-vd} \equiv {\cal I}_{l,2}^{vd-vd}$, we have that
\begin{equation}
\lim_{l\rightarrow \infty} C_{l,A}^{vd-vd} = \frac{2}{\pi} \int dk\;k^2 \; {\cal C}(k) \; ({\cal I}_l^{vd-vd})^2 \; \times 
\biggl( \frac{1}{4} - \frac{1}{4} - \frac{1}{4} + \frac{1}{4} \biggr) = 0.
\label{eq:limCBa}
\end{equation}

\subsubsection{The $C^{vd-vd}_B (\n_1,\n_2)$ contribution}

This case is again identical to the $C^{vv-dd}_B (\n_1,\n_2)$ contribution. The final result is
\[
C^{vd-vd}_A(\n_1,\n_2) = \sum_{l} \frac{2l+1}{4\pi} P_l(\n_1\cdot\n_2) C_{l,B}^{vd-vd} = 
\]
\begin{equation}
\phantom{xxxxxxxxxxxxx}
\sum_{l} \frac{2l+1}{4\pi} P_l(\n_1\cdot\n_2)\; \frac{2}{\pi}\int k^2dk\; (-{\cal D}(k)) \biggl[ {\cal I}_{l+1,1}^{vd-vd} 
{\cal I}_{l+1,2}^{vd-vd} \frac{l+1}{2l+1} +  {\cal I}_{l-1,1}^{vd-vd} 
{\cal I}_{l-1,2}^{vd-vd} \frac{l}{2l+1} 
\biggr] .
\label{eq:c12Cb1}
\end{equation}
This term gives most of the contribution of $C_l^{vd-vd}$ at high $l$-s, and it is {\em negative}. It is always below
the $vv-dd$ term, but it reduces the (positive) amplitude of that term. 
\begin{equation}
\lim_{l\rightarrow \infty} C_{l,B}^{vd-vd} = \frac{2}{\pi} \int dk\;k^2 \;(-{\cal D}(k)) \; ({\cal I}_l^{vv-dd})^2 = \lim_{l\rightarrow \infty} C_{l}^{vd-vd}.
\label{eq:limCBb}
\end{equation}

\label{lastpage}

\end{document}